\providecommand{\reserveinserts}[1]{}
\documentclass[AMA,STIX2COL,Linenumbersoff]{MRM}
\articletype{Research Article}%

\historydates{}
\doiheadtext{}
\usepackage{graphicx}
\usepackage{placeins}
\usepackage{dblfloatfix}
\usepackage{atbegshi}

\begin{document}

\title{Fast Reconstruction of Motion-Corrupted Data with Mobile-GRAPPA: 
\underline{Mo}tion and $\underline{\boldsymbol{\delta B_0}}$ 
\underline{I}nhomogeneity Correction \underline{L}everaging 
\underline{E}fficient 
\underline{GRAPPA}}


\author[1]{Yimeng Lin}{\orcid{0009-0002-6813-7102}}
\author[2]{Nan Wang}{\orcid{0000-0002-0081-873X}}
\author[1]{Daniel Abraham}{\orcid{0000-0003-4941-5840}}
\author[3]{Daniel Polak}{\orcid{0000-0001-9781-1528}}
\author[2]{Xiaozhi Cao}{\orcid{0000-0001-5095-648X}}
\author[2]{Aizada Nurdinova}{\orcid{0009-0000-9328-2155}}
\author[3]{Stephen Cauley}{}
\author[1,2]{Kawin Setsompop}{\orcid{0000-0002-4169-8447}}

\authormark{Lin \textsc{et al}}

\address[1]{\orgdiv{Department of Electrical Engineering}, \orgname{Stanford University}, \orgaddress{\state{California}, \country{United States}}}
\address[2]{\orgdiv{Department of Radiology}, \orgname{Stanford University}, \orgaddress{\state{California}, \country{United States}}}
\address[3]{\orgdiv{Siemens Medical Solutions USA}, \orgname{Inc.}, \orgaddress{\state{Malvern}, \country{United States}}}

\corres{Kawin Setsompop, Department of Radiology Stanford University, Stanford, CA, 94305. \email{kawins@stanford.edu}}

\finfo{This work was partially supported by R01MH116173; R01EB019437; R01HD114719; R21EB038677; Siemens Healthineers.}

\abstract[Abstract]{
\section{Purpose} To develop an accurate, fast, and generalizable motion-corrected MRI reconstruction pipeline that remains computationally efficient when incorporating hundreds to thousands of motion states per acquisition from high-temporal-resolution tracking.

\section{Methods} We propose a fast k-space preprocessing approach termed Mobile-GRAPPA that utilizes local GRAPPA operators to clean motion-corrupted data, after which the cleaned k-space data can be directly reconstructed using standard SENSE. Noise amplification and residual artifacts from Mobile-GRAPPA cleaning are characterized to evaluate reconstruction performance.

\section{Results} Characterization experiments using 3D MPRAGE at 3T and multi-echo 3D GRE at 3T and 7T with discretized motions demonstrates that the proposed reconstruction using standard SENSE with Mobile-GRAPPA achieve equivalent performance to a more computationally-demanding Aligned-SENSE that incorporates motion directly into the SENSE model. In highly motion-corrupted 1-mm 3D GRE at 3T with 1,620 motion-and-$\delta B_0$ estimates, and 1-mm 3D EPTI at 3T with 544 motion-and-$\delta B_0$ estimates, Mobile-GRAPPA enables SENSE reconstruction with negligible time penalty (reconstruction times: $\sim$15 s for GRE and $\sim$20 minutes for EPTI), whereas Aligned-SENSE is computationally prohibitive ($>$10 hours for GRE and $>$10 days for EPTI).

\section{Conclusion} Mobile-GRAPPA incorporates dense motion and $\delta B_0$ tracking into SENSE-based reconstruction with minimal time penalty, jointly correcting trajectory perturbations, coil reweighting, and $\delta B_0$-induced phase changes. It enables fast, high-quality reconstruction of strongly motion-corrupted data and can be used directly with standard SENSE as well as more advanced SENSE reconstructions with priors.
}


\wordcount{5143}

\makeatletter
\newif\ifdiscardedMRMblankpage
\discardedMRMblankpagefalse
\AtBeginShipout{%
  \ifdiscardedMRMblankpage\else
    \ifnum\value{page}=2\relax
      \AtBeginShipoutDiscard
      \global\discardedMRMblankpagetrue
      \global\advance\c@page by -1\relax
    \fi
  \fi
}
\makeatother

\maketitle

\section{Introduction}\label{sec1}

Motion has long been a major challenge in MRI \cite{zaitsev2015motion, godenschweger2016brainmoco, maclaren2013pmcreview} because it simultaneously (i) perturbs k-space sampling locations \cite{bammer2007agSense}, (ii) alters relative coil-sensitivity weighting \cite{aksoy2008coilSensitivityPI,luengviriya2010sensProfile, yarach2015coilSensitivityMiscalibration, farajidana2016coilMotion}, and (iii) induces spatially varying polarizing magnetic field perturbations ($\delta B_0$) \cite{liu2018headmotionB0,brackenier2022posedependentB0}. These effects lead to ghosting, blurring, and coil-weighting-related artifacts, and are especially severe in long-TE, $T_2^*$-weighted acquisitions where $\delta B_0$ can produce large image phase perturbations.

Over the past decade, motion-tracking technologies have advanced rapidly. 
External measurement systems
\cite{maclaren2012micromotion,vanniekerk2019plugplay,vanniekerk2019wrad,zaitsev2006opticalpmc,stucht2015highestrespmc,gretsch2020fatnavsMPT,jorge2018trackdots,laustsen2022eegmotion,berglund2021markerlessdwi,chen2023multishotmarkerless, vionnet2021jointfieldmotion, ludwig2021pilottonecine, speier2015ptnav, wilkinson2021pilottoneuhf, brackenier2024pilottonedmc,anand2024beatpilottone},
and navigator-based approaches
\cite{vanderkouwe2006cloverleaf,white2010promo,tisdall2012vnavigators,wallace2019fidnavs,gallichan2016fatnavs7t,polak2022samer,ulrich2024servonavigators,brackenier2024queen,wang2025smena,vanniekerk2024shortTRpmc} can now provide sub-second estimates of rigid-body motion and, in some cases, $\delta B_0$ with high precision, yielding hundreds to thousands of motion states during a single scan. Here, a motion state denotes a time interval over which pose and field perturbations are assumed constant. For example, a 1-mm whole-brain 3D multi-echo GRE acquisition with SMENA navigators \cite{wang2025smena} at 0.4~s resolution can produce on the order of 1{,}600 motion and $\delta B_0$ states over a 10-min scan.

However, high-temporal-resolution tracking introduces a new bottleneck: incorporating thousands of states into reconstruction while remaining accurate and computationally feasible. Several families of reconstruction strategies such as Augmented-SENSE \cite{bammer2007agSense, yarach2016distortionVar} and Aligned-SENSE \cite{corderogrande2016alignedMultishot, liu2020reducingMotionSensitivity} have been proposed. In these approaches, each motion state contributes a distinct encoding operator, and iterative reconstruction must repeatedly apply its forward and adjoint operators. As the number of states grows into the hundreds or thousands, the computational cost becomes prohibitive. Practical implementations therefore often resort to temporal down-sampling or clustering \cite{liu2020reducingMotionSensitivity, meng2025qsmB0}, discarding expensive-to-acquire information. This tension between dense motion information and reconstruction cost is most severe in long, high-resolution acquisitions where motion is common. It becomes more acute when the motion-free reconstruction is already computationally demanding, such as in 3D EPTI \cite{wang2019epti, wang2025septi}, due to the high-dimensional spatiotemporal encoding and accompanying subspace-based reconstruction \cite{liang2007spatiotemporal,dong2020eptiSubspace}. Explicitly incorporating hundreds of motion states into the EPTI forward operator pushes computation beyond practical limits \cite{dong2022motionCorrected3dEPTI}. Consequently, modern acquisitions can generate more motion and $\delta B_0$ information than current reconstructions can afford to use. 

To reduce this burden, a commonly used approximation assumes that coil sensitivity profiles move rigidly with the head and that $\delta B_0$ changes are negligible. Under these assumptions, a single encoding operator can be shared across all motion states, keeping the computational cost on the same order as the motion-free reconstruction \cite{xu2019mrfSlidingWindow,cruz2019mcmrf}. While this approximation can be adequate for small motion, mild field perturbations, and short-TE sequences, it fails under larger motion, substantial coil-weighting variation, or pronounced susceptibility effects, particularly at higher field strengths.

These observations motivate a different strategy: keep the reconstruction operator simple and shift motion and $\delta B_0$ handling into a one-time k-space preprocessing step. Prior work has shown that local GRAPPA kernels can interpolate missing k-space samples \cite{griswold2002grappa}, perform gridding of non-Cartesian data \cite{seiberlich2007grog}, and act as correction operators for eddy-current-induced phase errors \cite{hoge2016dpgrappa, wang2024fcg}. Here, we introduce Mobile-GRAPPA, a k-space framework that uses local GRAPPA operators to clean motion-corrupted k-space by correcting state-dependent coil reweighting and $\delta B_0$-induced phase, while simultaneously mapping the corrected data onto a prescribed Cartesian k-space grid for FFT-based downstream reconstruction \cite{lin2026mobilegrappa, lin2025motionigrog}. Building on recent work that jointly represents GRAPPA kernel families using a multilayer perceptron (MLP) \cite{abraham2023implicitgrappa}, we exploit the smooth variation of optimal kernels across nearby k-space locations and similar motion states to efficiently learn a continuous family of location- and state-dependent kernels with a MLP to clean motion-corrupted k-space data. After this one-time cleaning step, the data can be reconstructed using standard SENSE or more advanced pipelines, enabling fast, high quality reconstruction without modifying their forward operators. In extreme motion regimes, we further improve stability using a clustered variant that cleans to intermediate motion states before final reconstruction.

This work makes three contributions. First, we propose Mobile-GRAPPA as a plug-and-play k-space preprocessing module that enables fast and accurate reconstruction under motion and $\delta B_0$ variation. Second, we introduce an MLP-based parameterization of the Mobile-GRAPPA kernel family that exploits smooth structure across motion, $\delta B_0$, and k-space location. Third, we propose a clustering strategy for extreme motion to improve conditioning and stability.

\section{Theory}\label{sec2}

In this section, we first review Augmented-SENSE, approximate Augmented-SENSE, and Aligned-SENSE to establish the accuracy-efficiency trade-offs of existing motion-aware reconstructions and to motivate Mobile-GRAPPA. We then describe the proposed Mobile-GRAPPA framework and its MLP-based kernel representation.

\begin{figure*}[!t]
\centering
\includegraphics[width=\linewidth,height=0.65\textheight,keepaspectratio]{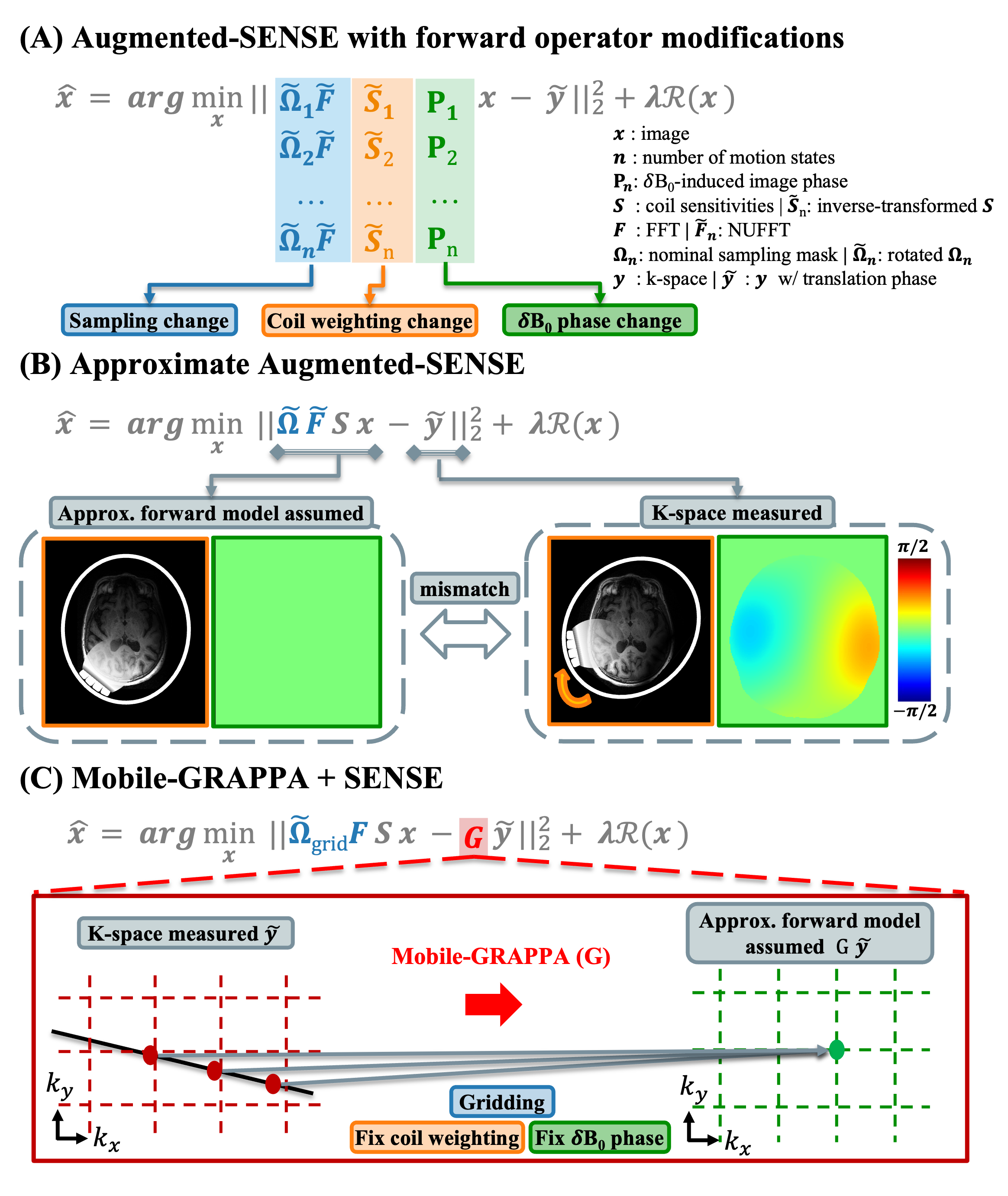}
\caption{
Overview of motion-aware reconstruction formulations and the motivation for Mobile-GRAPPA.
(A) Augmented-SENSE accounts for motion through state-dependent changes in sampling, coil weighting, and $\delta B_0$ phase.
(B) Approximate Augmented-SENSE reduces computational cost by simplifying the forward model, but this introduces mismatch between the assumed model and the measured k-space data.
(C) Mobile-GRAPPA addresses this mismatch by applying local k-space kernels $G$ to clean the motion-corrupted data and map them onto a prescribed Cartesian grid, after which reconstruction proceeds with standard SENSE.
}
\label{fig1}
\end{figure*}

\subsection{Augmented-SENSE}

We denote the complex transverse magnetization in a reference pose by $x(\mathbf{r})$ and the coil sensitivity maps in the same coordinate system by $S(\mathbf{r})$. The acquisition is partitioned into rigid-body motion states indexed by $n=1,\dots,N_{\mathrm{state}}$, where state $n$ is characterized by a rigid transform $T_n$ acting on spatial coordinates, with rotation $R_n$ and translation $\mathbf{c}_n$. When available, each state also has an associated image phase perturbation $\phi_n(\mathbf{r},t)$ induced by $\delta B_0$. We denote the measured multi-coil signals by $y_n(t)$ and the nominal trajectory sample by $\mathbf{k}_n(t)$.

Augmented-SENSE accounts for rigid motion by transforming the sampling locations and the coil sensitivities \cite{bammer2007agSense,yarach2016distortionVar}. As summarized in Figure~1A, motion introduces three state-dependent changes in the forward model: a change in sampling locations due to rotation, a change in coil weighting due to the altered pose relative to the coil array, and a change in image phase due to $\delta B_0$ perturbations.

First, the trajectory is inverse-rotated:
\begin{equation}
\tilde{\mathbf{k}}_n(t) = R_n^{-1}\mathbf{k}_n(t),
\end{equation}
which is generally not on the Cartesian grid. Therefore, Augmented-SENSE typically requires non-uniform Fourier transform (NUFFT) to model the Fourier encoding on $\tilde{\mathbf{k}}_n(t)$, even when the underlying acquisition was originally Cartesian.

Second, translations are modeled as a linear phase applied to the measured signals:
\begin{equation}
\tilde{y}_n(t) = y_n(t)\, e^{i 2\pi \mathbf{k}_n(t)\cdot\mathbf{c}_n}.
\label{eq:aug_trans_phase}
\end{equation}

Third, coil sensitivities are inverse-transformed into the pose of each state:
\begin{equation}
\tilde{S}_n(\mathbf{r}) = T_n^{-1}\!\left[S(\mathbf{r})\right].
\label{eq:aug_coil_transform}
\end{equation}

With these definitions, the Augmented-SENSE signal model becomes
\begin{equation}
\tilde{y}_n(t)
=
\int \big[x(\mathbf{r}) e^{i\phi_n(\mathbf{r},t)}\big]\tilde{S}_n(\mathbf{r}) \,
e^{-i 2\pi \tilde{\mathbf{k}}_n(t)\cdot\mathbf{r}} \,
d\mathbf{r}.
\label{eq:augmented_sense_cont}
\end{equation}

In operator form, corresponding to Figure~1A, let $\tilde F$ denote the NUFFT operator. Let $\tilde\Omega_n$ denote the sampling operator for motion state $n$, corresponding to sampling on $\tilde{\mathbf{k}}_n(t)$. Let $\tilde S_n$ denote the coil encoding operator formed from the inverse-transformed sensitivity maps. Let $P_n$ denote the image-domain phase modulation induced by $\delta B_0$ during state $n$.
The state $n$ encoding operator is
\begin{equation}
\tilde E_n = \tilde\Omega_n\, \tilde F\, \tilde S_n\, P_n,
\end{equation}
and Augmented-SENSE estimates image $x$ from k-space data $\tilde y_n$ by solving
\begin{equation}
\hat x
=
\arg\min_x
\sum_{n=1}^{N_{\mathrm{state}}}
\left\|
\tilde E_n x - \tilde y_n
\right\|_2^2
\;+\;
\lambda\,\mathcal{R}(x),
\label{eq:augmented_sense_obj}
\end{equation}
where $\lambda \ge 0$ controls the strength of regularization and $\mathcal{R}(\cdot)$ is a chosen image regularizer such as Tikhonov or a learned prior.

Iterative solvers require repeated application of all $\tilde E_n$ and $\tilde E_n^H$, so the per-iteration arithmetic cost scales approximately linearly with $N_{\mathrm{state}}$. Moreover, storing and regenerating the state-dependent components increases memory usage, so practical runtime can exceed the ideal linear scaling.

\subsection{Approximate Augmented-SENSE}
\label{sec:approx_aug_sense}

To reduce computational burden, Approximate Augmented-SENSE (Approx-Aug-SENSE) retains only the trajectory inverse-rotation and the translation phase term, while dropping explicit coil transformations and detailed $\delta B_0$ modeling \cite{xu2019mrfSlidingWindow,cruz2019mcmrf}. Under these approximations, $\tilde S_n(\mathbf{r})$ is replaced by a single reference sensitivity map $S(\mathbf{r})$, and the phase term $e^{i\phi_n(\mathbf{r},t)}$ is neglected:
\begin{equation}
\tilde{y}_n(t)
\approx
\int x(\mathbf{r}) \, S(\mathbf{r}) \,
e^{-i 2\pi \tilde{\mathbf{k}}_n(t)\cdot\mathbf{r}} \, d\mathbf{r},
\label{eq:approx_aug_cont}
\end{equation}
Equivalently, all motion states share a single encoding operator,
\begin{equation}
\tilde E = \tilde\Omega\, \tilde F\, S,
\end{equation}
where $\tilde\Omega$ aggregates the inverse-rotated sampling locations across states. Reconstruction then solves
\begin{equation}
\hat x = \arg\min_x  \left\| \tilde E x - \tilde y \right\|_2^2
\;+\;
\lambda\,\mathcal{R}(x)
\end{equation}
with $\tilde y$ denoting the aggregated measured signals after applying the translation phase term.

This approximation removes the explicit state dependence from the forward operator and therefore avoids the near-linear growth in per-iteration cost with the number of motion states. However, this efficiency comes from assuming that state-dependent coil reweighting and motion-induced $\delta B_0$ phase are negligible. Figure~1B illustrates the consequence of this simplification: the assumed forward model omits the pose-dependent coil weighting and phase variation present in the measured k-space data, creating a mismatch between model and data. As motion, TE and field strength increase, this mismatch can become a dominant source of reconstruction error even when the sampling change itself is modeled.

\subsection{Aligned-SENSE}

Aligned-SENSE provides a complementary formulation that emphasizes image-domain motion modeling \cite{corderogrande2016alignedMultishot,liu2020reducingMotionSensitivity}. Using the same notation as above, the measured signals for motion state $n$ at readout time $t$ are
\begin{equation}
y_n(t)
=
\int S(\mathbf{r}) \, T_n\!\big[x(\mathbf{r}) e^{i\phi_n(\mathbf{r},t)}\big]\,
e^{-i 2\pi \mathbf{k}_n(t)\cdot\mathbf{r}} \, d\mathbf{r}.
\label{eq:aligned_cont}
\end{equation}

Aligned-SENSE is particularly convenient for Cartesian acquisitions because it does not require trajectory modification and thus avoids replacing FFT-based operators with NUFFT, as in the Augmented-SENSE approach after inverse-rotating k-space samples. In operator form, let $F$ denote the FFT and $\Omega_n$ the nominal sampling mask for state $n$. The state $n$ encoding operator is
\begin{equation}
E_n = \Omega_n\, F\, S\, T_n\, P_n,
\end{equation}
and reconstruction solves
\begin{equation}
\hat x = \arg\min_x \sum_{n=1}^{N_{\mathrm{state}}} \left\| E_n x - y_n \right\|_2^2 
\;+\;
\lambda\,\mathcal{R}(x).
\end{equation}
As in Augmented-SENSE, iterative solvers require repeated application of every $E_n$ and $E_n^H$. Although the motion operator $T_n$ can be computed efficiently \cite{corderogrande2016alignedMultishot}, the total per-iteration cost still becomes prohibitive when hundreds to thousands of motion states are present.

\begin{figure*}[!t]
\centering
\includegraphics[width=0.95\textwidth,height=0.75\textheight,keepaspectratio]{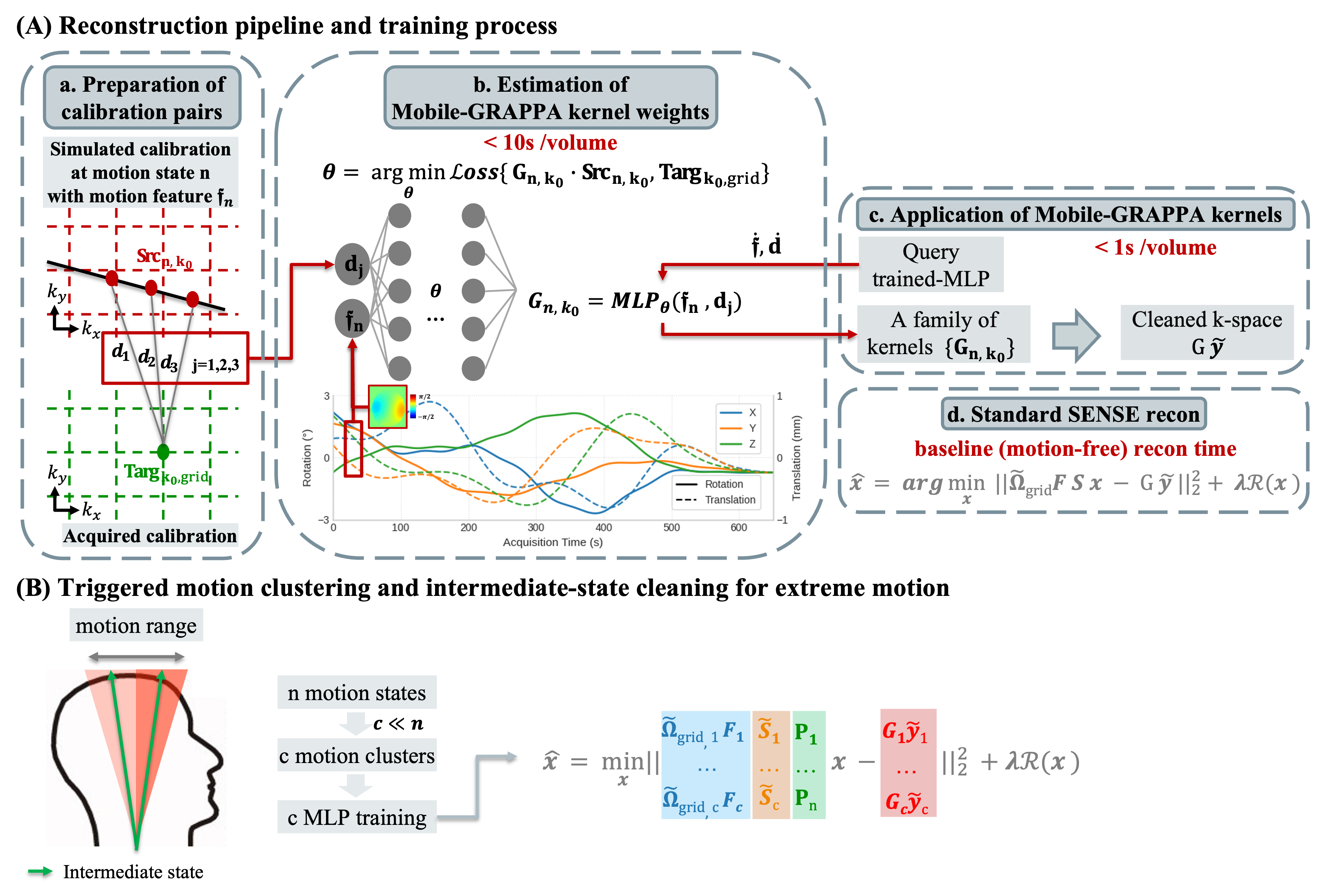}
\caption{
Mobile-GRAPPA reconstruction pipeline and clustering strategy.
(A) Reconstruction and training workflow.
(a) Preparation of calibration pairs from low-resolution fully sampled calibration data, where local source neighborhoods are extracted from simulated motion-corrupted k-space and paired with clean targets on the prescribed Cartesian grid.
(b) Estimation of Mobile-GRAPPA kernels using an MLP that maps motion-and-$\delta B_0$ features together with local source-to-target geometric information to a continuous family of kernels.
(c) Application of the trained kernels to the acquired motion-corrupted data to generate cleaned k-space on the target grid.
(d) Downstream reconstruction using the standard SENSE pipeline without motion-dependent forward-operator modification.
(B) Triggered clustering for extreme motion. When motion range is large, the original motion states are grouped into a smaller number of clusters, and Mobile-GRAPPA first cleans the data to nearby intermediate states before final reconstruction, improving conditioning and reducing noise amplification.
}
\label{fig2}
\end{figure*}

\subsection{Mobile-GRAPPA+SENSE}
\label{sec:mobile_grappa_pipeline}

This work aims to provide a generalizable reconstruction pipeline for motion-corrupted data that maintains high accuracy with minimal additional runtime. We build on the Approx-Aug-SENSE forward model (Section~\ref{sec:approx_aug_sense}), which achieves favorable computational scaling but, as illustrated by the mismatch in Figure~1B, no longer matches the acquired data when state-dependent coil reweighting and $\delta B_0$-induced phase are non-negligible. We therefore introduce Mobile-GRAPPA as a plug-and-play k-space preprocessing module that removes this mismatch in a one-time step so that the subsequent reconstruction can remain fast and unchanged.

Mobile-GRAPPA applies a family of local GRAPPA operators $\{G\}$ on the measured data with two coupled functions:
(i) removal of the mismatch between Approx-Aug-SENSE forward operator and the acquired data by correcting state-dependent coil reweighting and $\delta B_0$-induced image phase; and
(ii) replacement of expensive NUFFT operations with FFTs by mapping the corrected k-space data onto the prescribed $R_1$ Cartesian grid, where $R_1$ denotes the fully sampled grid used for the downstream reconstruction. This idea is summarized in Figure~1C: Mobile-GRAPPA transforms the motion-corrupted measurements into k-space data that better match the approximate forward model while simultaneously gridding them onto the target Cartesian locations.

The subsequent reconstruction solves
\begin{equation}
\hat x = \arg\min_x  \left\| \tilde\Omega_{\mathrm{grid}}\,  F\, S\, x - G\, \tilde y \right\|_2^2
\;+\;
\lambda\,\mathcal{R}(x),
\end{equation}
where $\tilde\Omega_{\mathrm{grid}}$ denotes $\tilde\Omega$ resampled onto the prescribed Cartesian $R_1$ grid.

Operationally, as shown in Figure~1C, we first apply the standard Approx-Aug-SENSE preprocessing for each motion state: inverse-rotate the trajectory and apply the translation phase to the measured k-space data. Let $\mathbf{k}_0$ denote a sample location on the inverse-rotated trajectory, and let $\mathbf{k}_{0,\mathrm{grid}}$ denote the nearest point to $\mathbf{k}_0$ on the prescribed Cartesian $R_1$ grid. Mobile-GRAPPA then uses a weighted combination of local multichannel samples around $\mathbf{k}_0$, acquired under the coil-weighting pattern associated with the head pose in state $n$ and the corresponding $\delta B_0$-induced phase, to produce a cleaned estimate at $\mathbf{k}_{0,\mathrm{grid}}$ with substantially reduced mismatch relative to the downstream reconstruction model. Aggregating these cleaned estimates across all motion states yields a multichannel dataset on the $R_1$ Cartesian grid that is substantially better matched to the Approx-Aug-SENSE forward model. The cleaned data can then be reconstructed using standard SENSE, or more advanced reconstructions with priors, without explicitly modifying the forward operator to incorporate motion or $\delta B_0$-induced phase.

\subsection{Mobile-GRAPPA training and MLP-based kernel representation}
\label{sec:mlp_training}
This subsection details the Mobile-GRAPPA preprocessing module introduced in Section~\ref{sec:mobile_grappa_pipeline}. Mobile-GRAPPA is trained using low-resolution fully sampled calibration data and then applied once to the motion-corrupted data to generate cleaned multichannel k-space on the $R_1$ Cartesian grid. As summarized in Figure~2A(a)--(c), the Mobile-GRAPPA procedure has three stages: (a) preparation of calibration pairs, (b) estimation of Mobile-GRAPPA kernel weights via an MLP, and (c) application of the trained kernels. The cleaned data are then passed to the downstream standard SENSE reconstruction shown in Figure~2A(d).

\subsubsection{Preparation of calibration pairs}

For each motion state we form training examples indexed by inverse-rotated trajectory sample $\mathbf{k}_0$ with associated target grid location $\mathbf{k}_{0,\mathrm{grid}}$. To form a source neighborhood for $\mathbf{k}_0$, we select the $J$ nearest samples to $\mathbf{k}_0$ along the readout dimension and denote their locations by $\{\mathbf{k}_j\}_{j=1}^J$.

This setup is illustrated in Figure~2A(a). The red points correspond to the local source neighborhood around $\mathbf{k}_0$ on the inverse-rotated trajectory, and the green point denotes the clean target on the prescribed Cartesian grid at $\mathbf{k}_{0,\mathrm{grid}}$. The relative offsets $\mathbf{d}_j = \mathbf{k}_j - \mathbf{k}_0$ provide the local geometric information relating the source samples to the target location.

Using fully sampled low-resolution calibration data, we extract the clean multichannel target at the corresponding grid location, denoted by $\mathbf{Targ}_{\mathbf{k}_{0,\mathrm{grid}}}$. Using rigid motion parameters and $\delta B_0$ estimates, we simulate calibration measurements with state-dependent coil weighting and $\delta B_0$-induced phase for each motion state and extract the corresponding source vectors $\mathbf{Src}_{n,\mathbf{k}_0}$. In this work, $\mathbf{Src}_{n,\mathbf{k}_0}$ stacks $J=5$ source samples per coil. This yields calibration pairs $\big(\mathbf{Src}_{n,\mathbf{k}_0}, \mathbf{Targ}_{\mathbf{k}_{0,\mathrm{grid}}}\big)$.

\subsubsection{Estimation of Mobile-GRAPPA kernel weights}

For each motion state $n$ and trajectory sample $\mathbf{k}_0$, we define a Mobile-GRAPPA kernel $\mathbf{G}_{n,\mathbf{k}_0}$ that maps the multichannel source neighborhood of $\mathbf{k}_0$ to the cleaned multichannel target at $\mathbf{k}_{0,\mathrm{grid}}$:
\begin{equation}
\mathbf{Targ}_{\mathbf{k}_{0,\mathrm{grid}}}
=
\mathbf{G}_{n,\mathbf{k}_0}\,
\mathbf{Src}_{n,\mathbf{k}_0}.
\label{eq:mobile_grappa_kernel_training}
\end{equation}
Therefore, $\mathbf{G}_{n,\mathbf{k}_0}$ maps from $(J\cdot N_{\mathrm{coil}})$ inputs to $N_{\mathrm{coil}}$ outputs.

A naive implementation would estimate and store a distinct kernel $\mathbf{G}_{n,\mathbf{k}_0}$ for every motion state and every sampled k-space location. Thus, for a single-echo acquisition, the total number of kernels scales as $N_{\mathrm{state}} \times N_k$, where $N_k$ is the number of acquired k-space samples. For a multi-echo acquisition, this scaling becomes $N_{\mathrm{state}} \times N_{\mathrm{echo}} \times N_k$. This quickly becomes impractical because the kernels vary jointly with motion state, echo, and sample location. In practice, however, kernel weights are strongly correlated across similar poses, nearby k-space locations, and adjacent echoes. We therefore represent the kernel family compactly using an MLP \cite{abraham2023implicitgrappa}.

We learn $\{\mathbf{G}_{n,\mathbf{k}_0}\}$ using an MLP trained on the calibration pairs, as illustrated in Figure~2A(b). The MLP takes motion-and-$\delta B_0$ features $\mathbf{f}_n$ together with the local source-to-target geometric information $\{\mathbf{d}_j\}$ and provides a nonlinear parameterization of the kernel family across motion states and k-space sample locations:
\begin{equation}
\mathbf{G}_{n,\mathbf{k}_0} = \mathcal{MLP}_\theta(\mathbf{f}_n, \{\mathbf{d}_j\}),
\end{equation}
where $\theta$ are network parameters.

Rather than providing full $\delta B_0$ maps to the network, we represent state-dependent $\delta B_0$ using spherical-harmonic coefficients \cite{stockmann2018shimming}, which provide a compact description of smooth field perturbations while avoiding high-dimensional image-domain inputs. In this work, the 3T $\delta B_0$ estimates from SMENA were modeled with second-order spherical harmonics, so all 9 normalized coefficients were used directly as features. In contrast, the 7T $\delta B_0$ estimates required higher-order models from more advanced estimation of st-SMENA \cite{wang2026stsmena}, typically sixth- to eighth-order, corresponding to 49--81 coefficients. We therefore applied singular value decomposition (SVD) across the coefficient vectors and retained enough components to preserve 99\% of the energy, which required only 6 SVD components to provide a compact representations in practice. More generally, for settings with higher-order $\delta B_0$ structure, we recommend fitting up to eighth-order spherical harmonics followed by SVD compression to 99\% retained energy.

During training, we stochastically sample trajectory locations $\mathbf{k}_0$ and motion states $n$. We use an $\ell_1$ data term to measure mismatch between Mobile-GRAPPA-cleaned k-space and the clean target, and optionally add $\ell_2$ regularization on the kernel weights:
\begin{equation}
\mathcal{L} = \sum_{n,\mathbf{k}_0}\left\| \mathbf{G}_{n,\mathbf{k}_0} \mathbf{Src}_{n,\mathbf{k}_0} - \mathbf{Targ}_{\mathbf{k}_{0,\mathrm{grid}}} \right\|_1
+ \lambda_G \left\| \mathbf{G}_{n,\mathbf{k}_0} \right\|_2^2,
\end{equation}
\begin{equation}
\theta
=
\arg\min_\theta
\mathcal{L}.
\end{equation}
As shown in Figure~2A(b), the training stage predicts a continuous family of kernels from the motion features and local offsets rather than fitting each kernel independently. In practice, the MLP converges quickly, with typical training times of $<10$ s per volume on a single NVIDIA A5000 GPU.

\subsubsection{Application of kernels}

After training, we query the MLP with motion-and-$\delta B_0$ features $\mathbf{f}_n$ and local source-to-target geometric information $\{\mathbf{d}_j\}$ for each trajectory sample $\mathbf{k}_0$ in the acquired motion-corrupted data to obtain $\{\mathbf{G}_{n,\mathbf{k}_0}\}$. As illustrated in Figure~2A(c), these kernels are then applied through \eqref{eq:mobile_grappa_kernel_training} to produce cleaned multichannel estimates at $\mathbf{k}_{0,\mathrm{grid}}$.

Aggregating these outputs across all trajectory samples yields the cleaned multichannel k-space dataset $G\tilde y$ on the prescribed Cartesian grid. The cleaned data are then passed to the downstream reconstruction in Figure~2A(d), which uses the same standard SENSE reconstruction described in Section~\ref{sec:mobile_grappa_pipeline}. In practice, this kernel-application and cleaning step is also efficient, with typical runtimes of $<1$ s per volume on a single NVIDIA A5000 GPU.

\subsection{Clustering for extreme motion}
\label{sec:clustering}

For moderate motion ranges, a single MLP can model kernel variation across all motion and $\delta B_0$ states. In extreme motion regimes or under strong pose-dependent coil sensitivity variation, performance can degrade for two related reasons. First, the mapping from motion-and-$\delta B_0$ features to optimal kernel weights can vary more rapidly across a large pose range, reducing interpolation reliability for a single MLP. Second, and often more limiting, the required one-step k-space cleaning can become poorly conditioned: when the coil encoding is insufficiently informative for the correction implied by large motion or strong $\delta B_0$, the learned linear operator may amplify noise.

To improve stability and reduce noise amplification during the Mobile-GRAPPA cleaning process, we reduce the required correction per application by cleaning the acquired data from each motion-and-$\delta B_0$ state to a nearby intermediate state rather than directly to the motion- and $\delta B_0$-free state. We trigger k-means clustering \cite{macqueen1967classification,arthur2006kmeanspp} when the range of rotations across states exceeds a preset threshold. For the 3T experiments in this work, the threshold was set to $10^\circ$. Motion states are clustered in rigid-body parameter space, and each cluster $m\in\{1,\dots,N_{\mathrm{c}}\}$ is assigned an intermediate representative pose $T_m$. As illustrated in Figure~2B, the green arrow denotes such an intermediate pose, which reduces the effective motion range that each Mobile-GRAPPA kernel set must correct. For each cluster, we train a cluster-specific MLP that predicts kernels using relative motion to intermediate pose $m$ and the corresponding $\delta B_0$ features.

The intermediate-state cleaning operation produces a cleaned estimate on the $R_1$ grid consistent with the intermediate pose:
\begin{equation}
\mathbf{Targ}_{m, \mathbf{k}_{0,\mathrm{grid}}}
=
\mathbf{G}_{m,\mathbf{k}_0}\,
\mathbf{Src}_{n,\mathbf{k}_0},
\label{eq:intermediate_cleaning}
\end{equation}
where $\mathbf{Targ}_{m, \mathbf{k}_{0,\mathrm{grid}}}$ denotes the desired multichannel sample at $\mathbf{k}_{0,\mathrm{grid}}$ with coil weighting and $ \delta B_0$-induced phase referenced to intermediate pose $m$.

After cleaning, we aggregate samples within each intermediate state to form $N_{\mathrm{c}}$ gridded datasets and perform reconstruction using a small set of discrete states rather than the full set of original motion states. The intermediate-state-$m$ encoding operator is:
\begin{equation}
\tilde E_m = \tilde\Omega_{\mathrm{grid},m}\,  F\, \tilde S_m\, P_m,
\end{equation}
 and the reconstruction solves:
\begin{equation}
\hat x = \arg\min_x \sum_{m=1}^{N_c} \left\| \tilde E_m x - G_m\, \tilde y_m \right\|_2^2
\;+\;
\lambda\,\mathcal{R}(x).
\end{equation}

This strategy avoids overly aggressive single-step corrections in very large-motion cases, improves conditioning, and maintains substantially lower reconstruction overhead than fully state-resolved formulations such as Aligned-SENSE or Augmented-SENSE. As summarized in Figure~2B, clustering compresses $N_{\mathrm{state}}$ motion states into a much smaller number $N_{\mathrm{c}} \ll N_{\mathrm{state}}$ of intermediate states. The downstream reconstruction therefore scales primarily with $N_{\mathrm{c}}$ rather than $N_{\mathrm{state}}$. This benefit comes at the cost of training $N_{\mathrm{c}}$ separate cluster-specific MLPs, introducing an additional one-time preprocessing cost that also scales with $N_{\mathrm{c}}$. Thus, compared with standard Mobile-GRAPPA + SENSE, the clustered Mobile-GRAPPA pipeline increases both preprocessing and reconstruction cost, but compared with fully state-resolved motion-aware reconstructions, it offers a substantially more favorable scaling in extreme-motion regimes.
  
\section{Methods}\label{sec3}

\begin{table*}[t]%
\caption{In vivo experiment protocols.\label{tab:invivo}}
\centering
\footnotesize
\setlength{\tabcolsep}{6pt}
{\renewcommand{\arraystretch}{1.15}
\newcolumntype{C}[1]{>{\centering\arraybackslash}p{#1}}
\begin{tabularx}{\textwidth}{
C{0.4cm} 
C{2.1cm}  
C{0.6cm}  
C{1.8cm}  
C{2.5cm}  
C{1.3cm}  
C{2.3cm}  
C{1.7cm}  
C{1.5cm}  
}
\toprule
\textbf{Exp.} &
\textbf{Sequence} &
\textbf{Field} &
\textbf{FOV (mm$^3$)} &
\textbf{Resolution (mm$^3$)} &
\textbf{TA (min)} &
\textbf{TE/TI/TR (ms)} &
\textbf{\# of Echoes} &
\textbf{\# of State} \\
\midrule
1 & MPRAGE         & 3T & $256\times256\times192$ & $1.26\times1.0\times1.0$ & 2.5 /pose & 2.19/1000/2300 & 1 & 12 \\
2 & multi-echo GRE & 3T & $240\times240\times216$ & $1.0\times1.0\times1.0$  & 10       & 10.3-43.1/-/50 & 14 & 12 \\
3 & multi-echo GRE & 7T & $240\times240\times216$ & $1.0\times1.0\times1.0$  & 10       & 6.2-30.3/-/50 & 8 & 6 \\
4 & multi-echo GRE & 3T & $240\times240\times216$ & $1.0\times1.0\times1.0$  & 10       & 6-29.2/-/50 & 8 & 1620 \\
5 & EPTI           & 3T & $240\times240\times216$ & $1.0\times1.0\times1.0$  & 2        & 10.2-60.7/-/50 & 48 & 544 \\
6 & EPTI           & 3T & $240\times240\times216$ & $1.0\times1.0\times1.0$  & 2        & 10.2-60.7/-/50 & 48 & 544 \\

\bottomrule
\end{tabularx}
}
\end{table*}

We performed six in vivo experiments to evaluate Mobile-GRAPPA for removing motion-induced k-space inconsistencies while preserving the simplicity of standard downstream reconstructions. Across experiments, Mobile-GRAPPA used the same classes of inputs: fully sampled low-resolution calibration data, motion and $\delta B_0$ estimates, the k-space sampling trajectory, and motion-corrupted k-space data. In experiments involving dynamic tracking, motion and $\delta B_0$ estimates were obtained using SMENA \cite{wang2025smena} or st-SMENA \cite{wang2026stsmena}, with experiment-specific temporal resolution described below. All in vivo studies were approved by the institutional review board, and written informed consent was obtained from all subjects prior to scanning.

Experiments~1--3 evaluate Mobile-GRAPPA accuracy in data with limited numbers of discrete motion and/or $\delta B_0$ states using Aligned-SENSE as a reference. Experiments~4--6 evaluate correction performance and computational feasibility in cases with large numbers of motion and $\delta B_0$ states, where Aligned-SENSE is prohibitively expensive. Table~\ref{tab:invivo} summarizes the experiments.

\paragraph{Experiment 1: coil reweighting correction under discretized static poses (3D MPRAGE, 3T).} 
Experiment~1 used 3D MPRAGE to isolate the effect of motion-induced coil weighting change, while minimizing $\delta B_0$ phase accrual because of the short TE. Data were acquired on a 3T MR scanner (MAGNETOM Vida; Siemens Healthineers, Forchheim, Germany) using a 20-channel head coil. Acceleration was $R=4$ using $2\times2$ CAIPI undersampling pattern \cite{breuer2005caipirinha}. Five separate scans were performed at different static head poses and retrospectively combined by mixing k-space segments across poses to form a synthesized motion-corrupted dataset with 12 motion states, with three of the five acuqired poses each contributing to three motion states, one contributing to two motion states and one contributing to one motion states. Rigid motion parameters were estimated by registering each pose image to a reference pose. Mobile-GRAPPA cleaned the mixed-pose k-space to the reference pose, after which the cleaned data were reconstructed using standard SENSE (Mobile-GRAPPA + SENSE). Reconstructions were compared against no correction, Approx-Aug-SENSE and Aligned-SENSE.

\paragraph{Experiment 2: $\delta B_0$-induced phase correction under step motion (multi-echo 3D GRE, 3T).}
Experiment~2 used multi-echo 3D GRE to evaluate Mobile-GRAPPA's ability to correct both motion and $\delta B_0$-induced phase and to assess its robustness across echo times. Data were acquired on a 3T MR scanner (GE UHP; GE Healthcare, Milwaukee, WI, USA) using a 32-channel head coil. Acceleration was $R=4$ using $2\times2$ uniform undersampling pattern. A healthy volunteer was instructed to perform 12 step motions during a single acquisition. Mobile-GRAPPA cleaning used motion and $\delta B_0$ estimates from SMENA navigators every 50~s. Because the $\delta B_0$-induced phase evolves with echo time, a separate set of Mobile-GRAPPA kernels was trained and applied for each echo. Mobile-GRAPPA was evaluated in three configurations: coil-weighting-only correction, $\delta B_0$-only correction, and joint correction. The cleaned data were reconstructed using standard SENSE. Reconstructions were compared against no correction, Approx-Aug-SENSE and Aligned-SENSE.

\paragraph{Experiment 3: high field robustness under step motion (multi-echo 3D GRE, 7T).}
Experiment~3 repeated the multi-echo 3D GRE characterization at 7T to test robustness under amplified high-field effects, where state-dependent coil sensitivity variation and $\delta B_0$ perturbations are more pronounced \cite{wiesinger2004piFieldStrength,stockmann2018shimming,liu2018headmotionB0}. Data were acquired on a 7T MR scanner (MAGNETOM Terra.X; Siemens Healthineers, Forchheim, Germany) using a 32-channel head coil. A healthy volunteer performed 6 step motion during the scan. For this particular acquisition, Motion and $\delta B_0$ were tracked densely by st-SMENA navigators every 0.4~s. For evaluation, we applied k-means clustering in rigid motion parameter space with associated $\delta B_0$ features to compress the motion information into 12 representative motion states that can well approximate the 6 stationary poses from the 5 step motions along with the transition poses. This kept the Aligned-SENSE reference reconstruction tractable. Mobile-GRAPPA + SENSE and comparisons against no correction, Approx-Aug-SENSE, and Aligned-SENSE were performed as in Experiment~2.

\paragraph{Experiment 4: feasibility under continuous free motion and dense tracking (multi-echo 3D GRE, 3T).}
Experiment~4 evaluates Mobile-GRAPPA's scalability with high-temporal-resolution tracking using the same protocol and scanner as in Experiment~2. The subject moved freely throughout the 10-min scan while SMENA provided motion and $\delta B_0$ estimates every 0.4~s, yielding $N_{\text{state}}=1{,}620$ motion-and-$\delta B_0$ states. Mobile-GRAPPA cleaned the k-space using all tracked states and the cleaned data were reconstructed using standard SENSE. Aligned-SENSE was evaluated only with temporally sub-sampled or compressed state sets due to prohibitive runtime at $N_{\text{state}}=1{,}620$.

\paragraph{Experiment 5: feasibility when motion-free reconstruction is computationally expensive (3D EPTI, 3T).}
Experiment~5 tested Mobile-GRAPPA on the same scanner as Experiment~2 but with 1 mm isotropic 3D Echo Planer Time-resolved Imaging (EPTI) \cite{wang2019epti, wang2025septi}. EPTI performs efficient spatiotemporal $k$-$t$ sampling with subspace reconstruction to recover distortion- and blurring-free time-resolved $T_2^\ast$-weighted images, making even the motion-free reconstruction computationally demanding \cite{dong2020eptiSubspace}. The subject moved freely during the 2-min acquisition while SMENA provided motion and $\delta B_0$ estimates every 0.2~s, yielding $N_{\text{state}}=544$ tracked states. Mobile-GRAPPA cleaned the k-space using all tracked states and the cleaned data were reconstructed using the standard EPTI subspace reconstruction pipeline. Explicit motion modeling with an Aligned-SENSE-style EPTI forward operator was evaluated only using temporally sub-sampled or compressed state sets due to prohibitive runtime scaling.

\paragraph{Experiment 6: feasibility under deep breath and dense tracking (3D EPTI, 3T).}
Experiment~6 isolates respiration-driven $\delta B_0$ perturbations by instructing the subject to minimize head motion and perform deep breathing. Acquisition and reconstruction settings were matched to Experiment~5 to attribute differences primarily to time-varying $\delta B_0$.

\paragraph{Quantitative evaluation.}
Residual artifacts were assessed using $5\times$-scaled error maps and normalized root-mean-square error (NRMSE) relative to the corresponding Aligned-SENSE reconstruction.

Noise amplification introduced by Mobile-GRAPPA was characterized using pseudo-replica analysis \cite{robson2008snrgfactor} with 200 noise realizations in Experiments~2 and~3. The reported g-factor reflects noise propagation through the Mobile-GRAPPA cleaning operator and not the parallel-imaging g-factor of the downstream SENSE reconstruction. Specifically, the g-factor was computed by comparing the noise characteristics of i) Mobile-GRAPPA + SENSE and ii) Aligned-SENSE. The acquired k-space data were first prewhitened using the coil-noise covariance matrix, after which 200 pseudo-replicas were generated by adding i.i.d. complex Gaussian noise to the prewhitened k-space for each reconstruction. The resulting $1/$g-factor maps therefore quantify the additional noise amplification attributable to Mobile-GRAPPA cleaning relative to the Aligned-SENSE reference.

\section{Results}\label{sec4}

\subsection{Discretized-state characterization (Experiments~1--3)}

\begin{figure*}[!t]
\centering
\includegraphics[width=\linewidth,height=0.58\textheight,keepaspectratio]{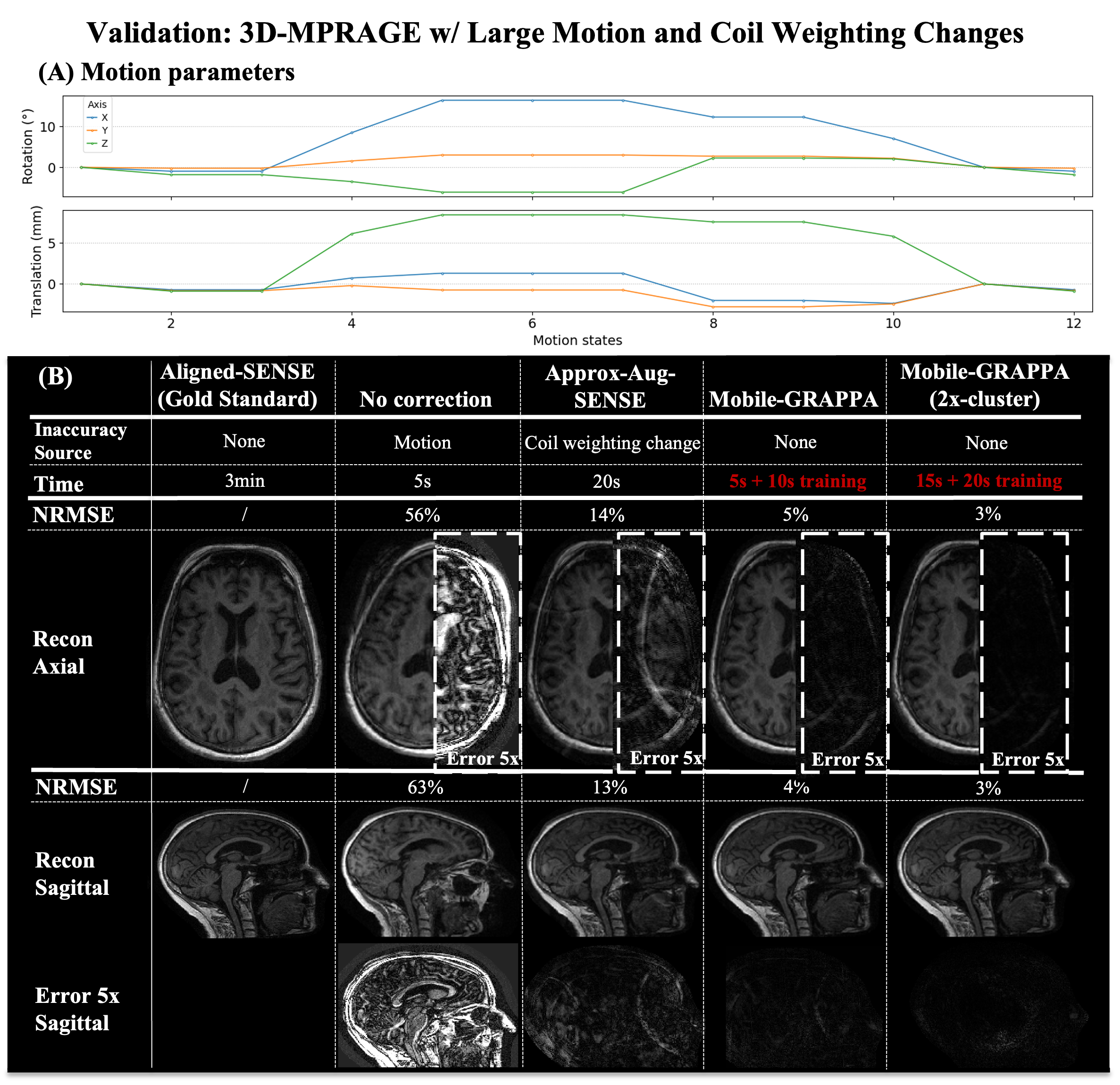}
\caption{
Experiment~1: synthesized 3D MPRAGE at 3T with large discrete motion and coil-weighting variation.
(A) Six rigid-body motion parameters across the 12 motion states.
(B) Reconstruction comparison. Column 1: Aligned-SENSE reference. Column 2: no correction. Column 3: Approx-Aug-SENSE. Column 4: Mobile-GRAPPA + SENSE. Column 5: Mobile-GRAPPA + SENSE with two-group clustering. Axial and sagittal reconstructions are shown together with $5\times$ error maps relative to Aligned-SENSE. Mobile-GRAPPA substantially reduces motion-related inconsistencies and approaches the Aligned-SENSE reference while maintaining near-baseline reconstruction time.
}
\label{fig3}
\end{figure*}

In the discretized MPRAGE experiment, Mobile-GRAPPA effectively corrected large state-dependent inconsistencies arising from very large motions, exceeding 10 degrees/mm (Figure~3A). The no-correction reconstruction exhibits severe motion artifacts, while Approx-Aug-SENSE reduces these artifacts but still leaves errors consistent with pose-dependent coil reweighting. Mobile-GRAPPA substantially suppresses these inconsistencies and approaches the Aligned-SENSE reference in both axial and sagittal views, with little remaining structure in the $5\times$ error maps. Quantitatively, NRMSE decreases from 56--63\% (no correction) to 4--5\% with Mobile-GRAPPA and further to 3\% with two-group clustering. The single-group Mobile-GRAPPA reconstruction incurs only a small fixed training overhead relative to baseline reconstruction, whereas the two-group clustered version increases both training and reconstruction time by approximately $2\times$.

\begin{figure*}[!t]
\centering
\includegraphics[width=\linewidth,height=0.7\textheight,keepaspectratio]{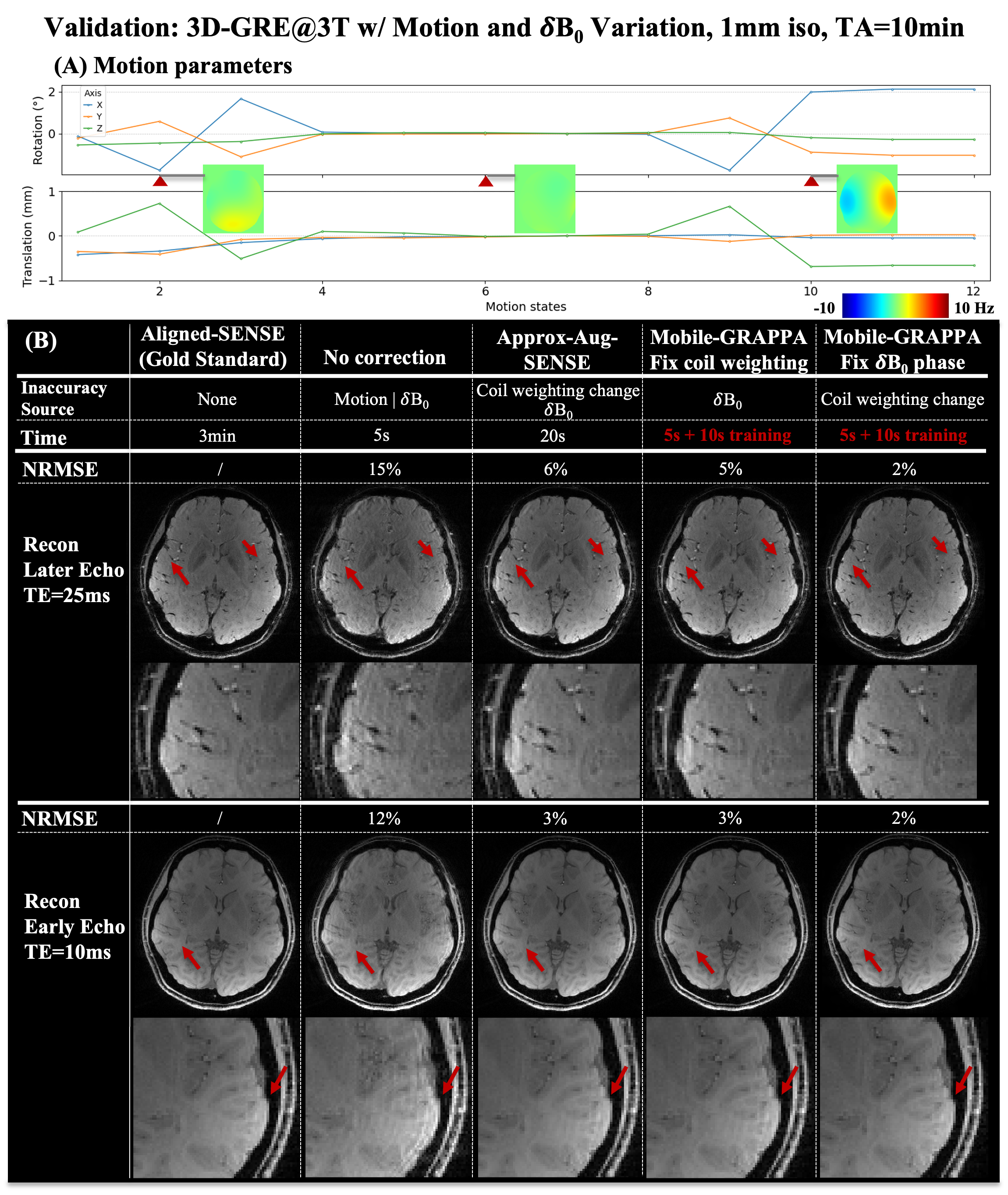}
\caption{
Experiment~2: step-motion multi-echo 3D GRE at 3T with motion and $\delta B_0$ variation.
(A) Six rigid-body motion parameters across the 12 motion states, with representative $\delta B_0$ maps shown for selected states.
(B) Reconstruction comparison at an early echo (TE = 10 ms) and a later echo (TE = 25 ms). Column 1: Aligned-SENSE reference. Column 2: no correction. Column 3: Approx-Aug-SENSE. Column 4: Mobile-GRAPPA correcting coil weighting only. Column 5: Mobile-GRAPPA correcting $\delta B_0$ phase only. Red arrows indicate representative residual artifacts. The results show that coil-weighting correction and $\delta B_0$ correction address complementary error sources, with $\delta B_0$ effects becoming more prominent at longer TE.
}
\label{fig4}
\end{figure*}

The step-motion multi-echo GRE experiment at 3T separates the roles of coil reweighting and $\delta B_0$ phase, and also shows that phase-related errors become more visible at longer echo time. Although Approx-Aug-SENSE removes the gross motion artifacts, residual inconsistencies remain, especially at the later echo (Figure~4B). At 3T, correcting coil weighting alone provides only modest additional improvement beyond Approx-Aug-SENSE, indicating that pose-dependent coil mismatch is a secondary error source in this setting. In contrast, $\delta B_0$-only Mobile-GRAPPA more effectively suppresses the residual artifacts at longer TE, consistent with increased phase accrual.

\begin{figure*}[!t]
\centering
\includegraphics[width=\linewidth,height=0.64\textheight,keepaspectratio]{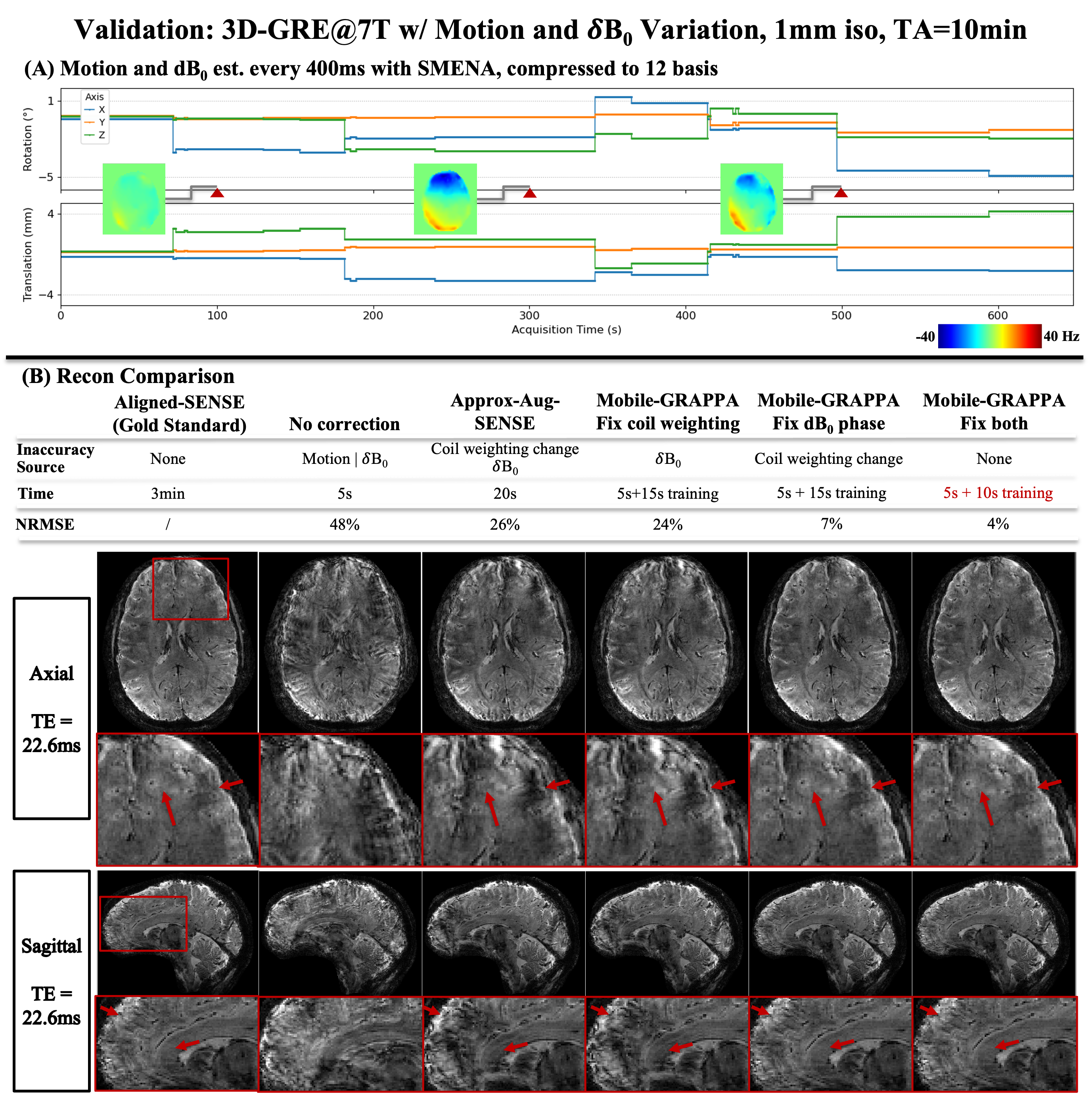}
\caption{
Experiment~3: step-motion multi-echo 3D GRE at 7T with stronger motion- and susceptibility-related effects.
(A) Motion and $\delta B_0$ estimates tracked by SMENA every 400 ms and compressed to 12 representative states for tractable Aligned-SENSE reconstruction.
(B) Reconstruction comparison at TE = 22.6 ms. Column 1: Aligned-SENSE reference. Column 2: no correction. Column 3: Approx-Aug-SENSE. Column 4: Mobile-GRAPPA correcting coil weighting only. Column 5: Mobile-GRAPPA correcting $\delta B_0$ phase only. Column 6: Mobile-GRAPPA jointly correcting both coil weighting and $\delta B_0$. Zoomed axial and sagittal views highlight residual artifacts. At 7T, joint correction is required to approach the Aligned-SENSE reference.
}
\label{fig5}
\end{figure*}

Figure~5 demonstrates that ultra-high-field motion correction requires joint handling of coil reweighting and $\delta B_0$ phase. At 7T, Approx-Aug-SENSE correction leaves pronounced residual artifacts, consistent with stronger pose-dependent coil weighting variation and increased susceptibility-driven field perturbations (Figure~5B). Coil-only and $\delta B_0$-only Mobile-GRAPPA remove their targeted error sources but leave complementary residual artifacts, whereas joint correction produces the cleanest reconstructions in both axial and sagittal views. These results indicate that both correction components are required to achieve near-reference performance at ultra high field under motion.

\begin{figure*}[!t]
\centering
\includegraphics[width=\linewidth,height=0.54\textheight,keepaspectratio]{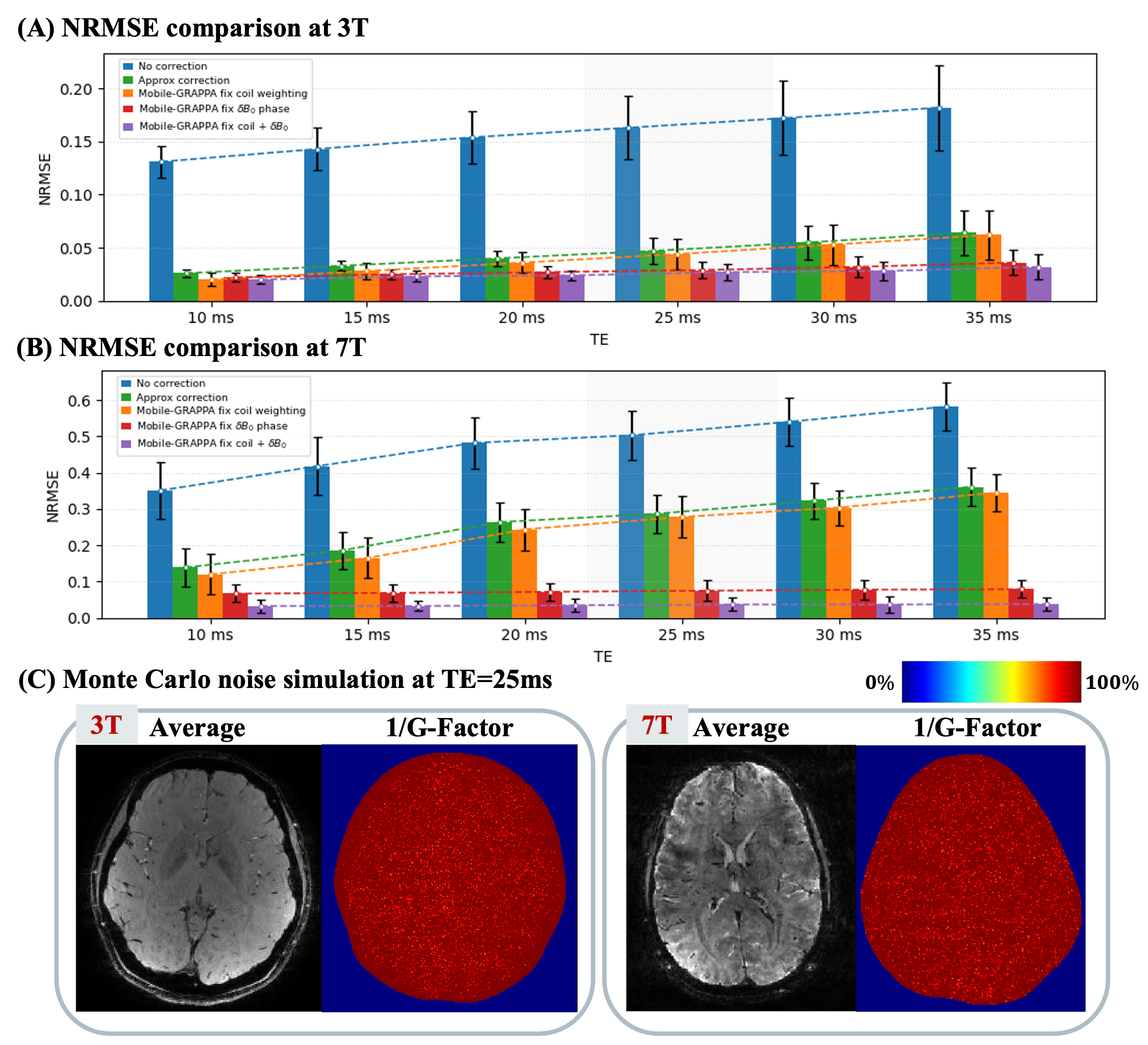}
\caption{
Quantitative evaluation for Experiments~2--3.
(A) Mean whole-brain NRMSE versus echo time at 3T.
(B) Mean whole-brain NRMSE versus echo time at 7T.
Error bars indicate the standard deviation across axial slices.
(C) Pseudo-replica noise analysis at TE = 25 ms, shown as average reconstructed images and empirical $1/g$-factor maps for 3T and 7T, with Aligned-SENSE used as the reference. Mobile-GRAPPA reduces reconstruction error across echoes without measurable additional noise amplification relative to Aligned-SENSE.
}
\label{fig6}
\end{figure*}

Figure~6 quantifies reconstruction error and noise behavior across echo time and field strength for Experiments~2--3. At 3T (Figure~6A), NRMSE increases with TE for no correction and for Approx-Aug-SENSE, while $\delta B_0$-corrected Mobile-GRAPPA reduces error across all echoes. The gap between the $\delta B_0$-corrected and non-$\delta B_0$-corrected variants widens with TE, indicating that field-related phase becomes increasingly important at longer echo times. In contrast, the difference between joint correction and $\delta B_0$-only correction remains relatively small at 3T, suggesting that explicit coil-weighting correction has a limited additional effect in this regime. 
At 7T (Figure~6B), both correction terms become more important: coil-weighting correction provides a clearer benefit than at 3T, and the contribution of $\delta B_0$ correction is markedly larger, particularly at longer TE. Together, these results support the view that field-related errors dominate at long TE, while the importance of coil-weighting correction grows with field strength. Pseudo-replica analysis at TE=25\,ms does not indicate measurable additional noise amplification attributable to Mobile-GRAPPA cleaning, as reflected by near-100\% $1/G$-factor maps when referenced to Aligned-SENSE (Figure~6C).

\subsection{Dense tracking: feasibility and runtime scaling (Experiments~4--6)}

\begin{figure*}[!t]
\centering
\includegraphics[width=\linewidth,height=0.81\textheight,keepaspectratio]{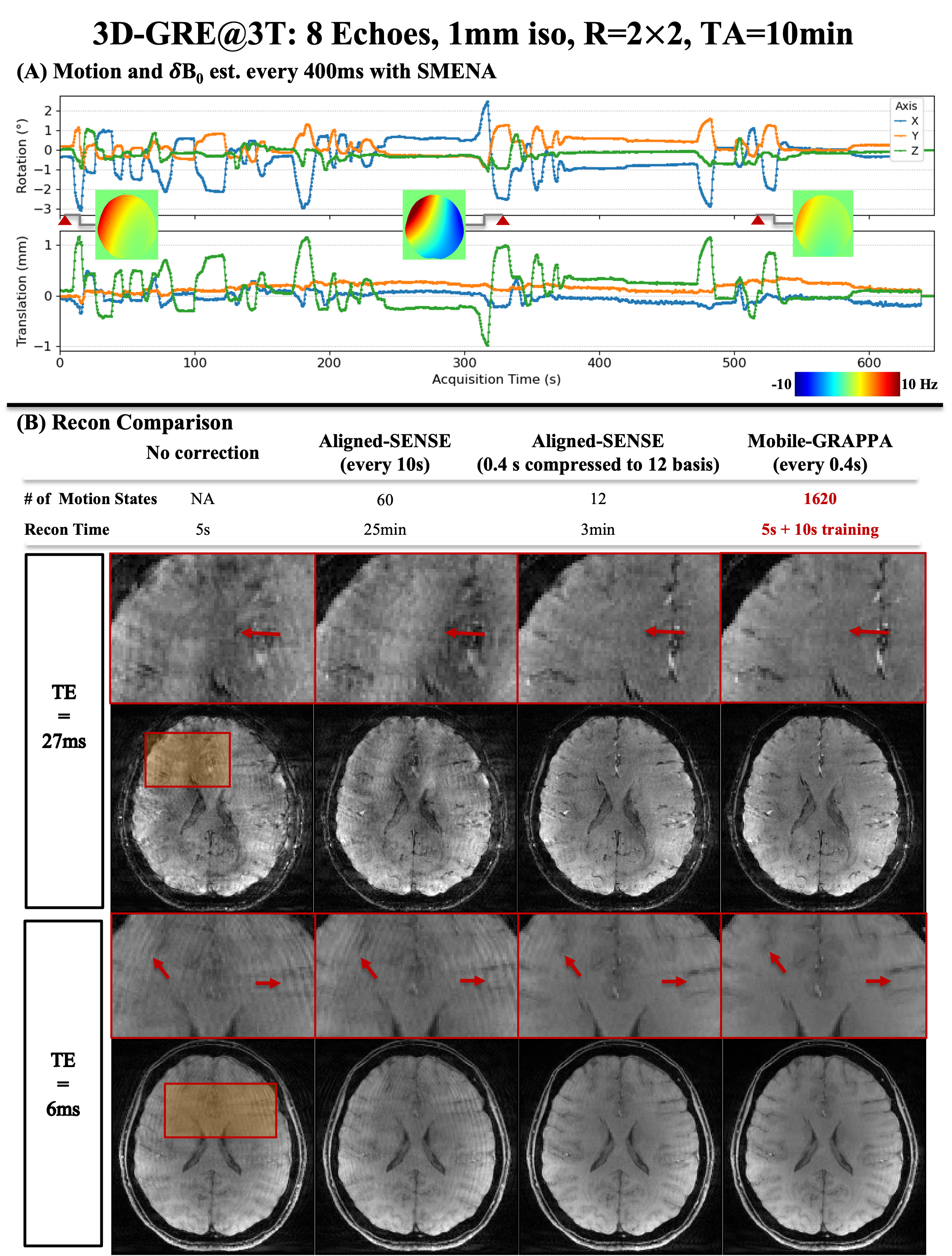}
\caption{
Experiment~4: dense motion and $\delta B_0$ tracking in 1-mm isotropic multi-echo 3D GRE at 3T.
(A) Motion and $\delta B_0$ estimates obtained every 400 ms using SMENA.
(B) Reconstruction comparison in the free-motion scan. Column 1: no correction. Column 2: Aligned-SENSE using temporally downsampled motion updates (every 10 s). Column 3: Aligned-SENSE using 12 clustered states derived from the dense motion estimates. Column 4: Mobile-GRAPPA using all 1,620 tracked motion-and-$\delta B_0$ states. Comparisons are shown for an early echo (TE = 6 ms) and a later echo (TE = 27 ms). Mobile-GRAPPA preserves dense motion information while maintaining near-baseline reconstruction time plus a small one-time training overhead.
}
\label{fig7}
\end{figure*}
The free-motion 3T GRE experiment highlights the scaling limitation of explicit state-dependent reconstruction under dense motion and $\delta B_0$ tracking. With 1{,}620 tracked motion-and-$\delta B_0$ states, Aligned-SENSE-style baselines require temporal down-sampling or state compression to remain computationally feasible, leading to a visible quality--runtime trade-off. Mobile-GRAPPA uses all 1{,}620 tracked states for a one-time k-space cleaning step and then preserves the standard SENSE reconstruction complexity (Figure~7). This enables consistent artifact suppression at both an early echo (TE = 6 ms), where motion-induced ringing is prominent, and a later echo (TE = 27 ms), where $\delta B_0$-related dephasing becomes more visible, while maintaining baseline-like reconstruction time with only a small fixed training overhead.

\begin{figure*}[!t]
\centering
\includegraphics[width=\linewidth,height=0.81\textheight,keepaspectratio]{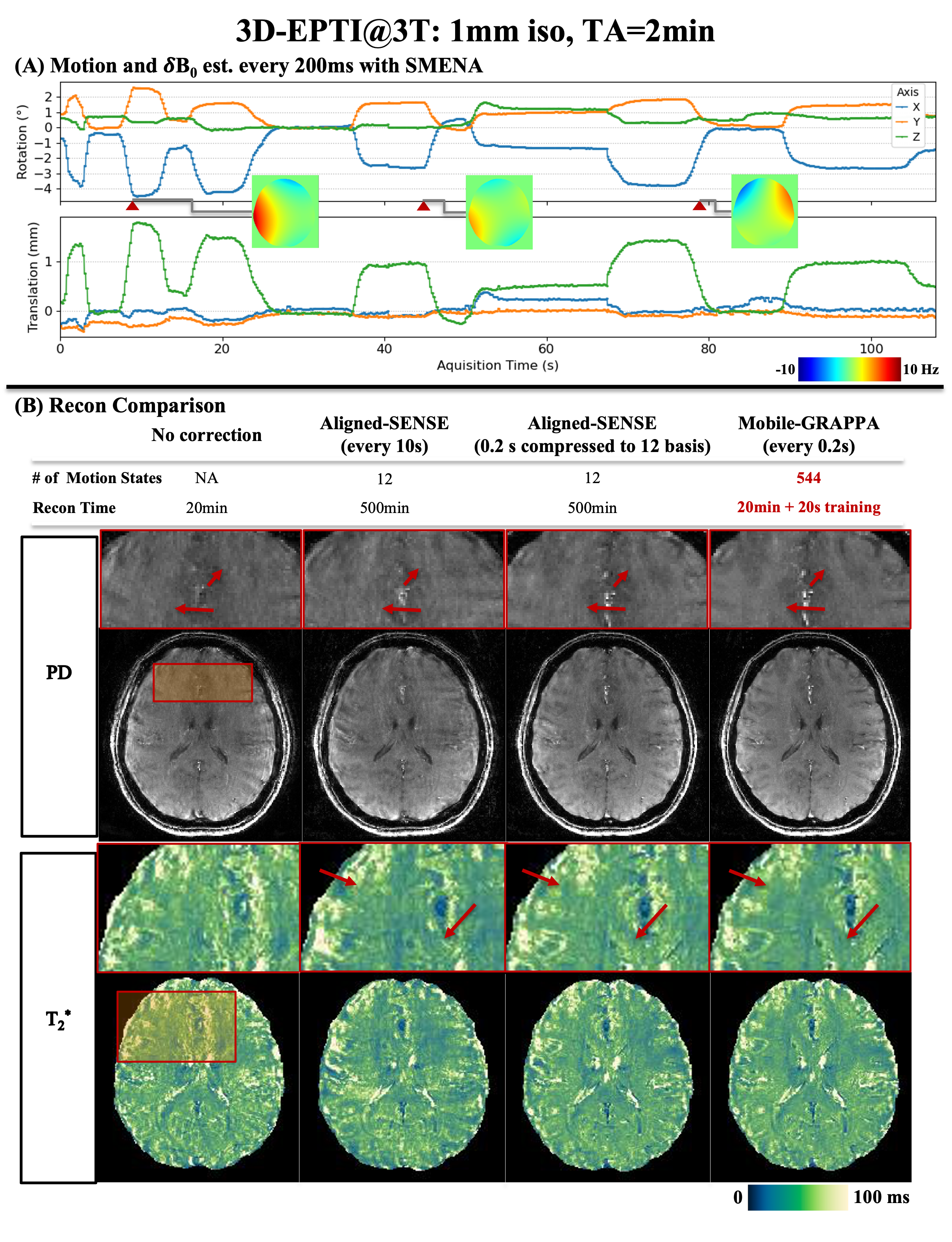}
\caption{
Experiment~5: 3D-EPTI at 3T with free motion and dense tracking.
(A) Motion and representative $\delta B_0$ maps estimated every 200 ms using SMENA.
(B) Reconstruction comparison for proton-density (PD) and $T_2^\ast$ outputs. Column 1: no correction. Column 2: Aligned-SENSE using temporally downsampled motion updates (every 10 s). Column 3: Aligned-SENSE using 12 clustered states derived from the dense motion estimates. Column 4: Mobile-GRAPPA using all 544 tracked motion-and-$\delta B_0$ states. Red arrows indicate representative residual artifacts. Mobile-GRAPPA improves image and parametric-map quality while preserving the baseline EPTI reconstruction pipeline and avoiding the prohibitive runtime of explicitly state-dependent reconstruction.
}
\label{fig8}
\end{figure*}

The same advantage becomes even more important for EPTI, where the motion-free reconstruction is already computationally expensive. In the 3D EPTI experiment with 544 tracked states, explicit state-dependent modeling remains prohibitively slow even after aggressive state compression, whereas Mobile-GRAPPA adds only a short one-time cleaning step and leaves the baseline subspace EPTI reconstruction unchanged (Figure~8). The resulting PD and $T_2^\ast$ outputs show visibly reduced motion-related artifacts in the highlighted regions, while the total runtime remains dominated by the original EPTI subspace reconstruction pipeline rather than the motion handling.

\begin{figure*}[!t]
\centering
\includegraphics[width=\linewidth,height=0.81\textheight,keepaspectratio]{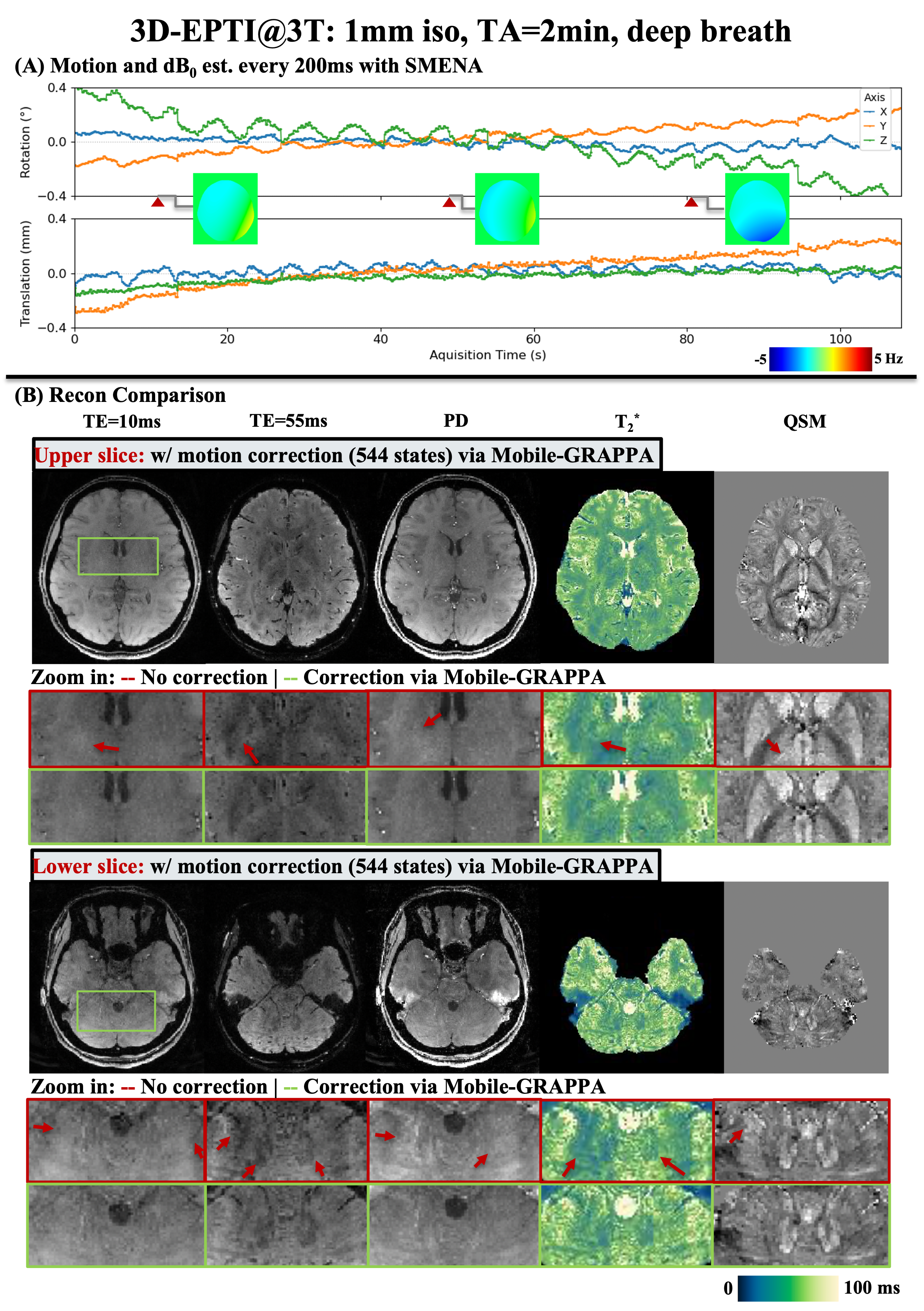}
\caption{
Experiment~6: 3D-EPTI at 3T under deep breathing with minimal intended rigid motion.
(A) Motion traces and representative $\delta B_0$ maps estimated every 200 ms using SMENA.
(B) Comparison of reconstructed multi-echo images, PD, $T_2^\ast$, and QSM \cite{langkammer2012qsmbrainiron} with and without Mobile-GRAPPA correction using all 544 tracked states. Red boxes indicate zoomed regions, and arrows highlight representative breathing-related artifacts that are reduced after Mobile-GRAPPA cleaning. The results show that Mobile-GRAPPA remains effective in a $\delta B_0$-dominant regime and improves both image quality and derived quantitative contrasts without modifying the downstream EPTI reconstruction.
}
\label{fig9}
\end{figure*}

A similar benefit is observed in the respiration-dominant experiment. Under deep breathing with minimal intended rigid motion, Mobile-GRAPPA reduces breathing-related artifacts in the multi-echo images and improves the derived PD, $T_2^\ast$, and QSM contrasts, particularly in susceptibility-sensitive regions (Figure~9B). This result indicates that the proposed one-time k-space cleaning can compensate time-varying $\delta B_0$ dynamics even when rigid motion is not the dominant source of corruption, again without modifying the downstream EPTI reconstruction.

\section{Discussion}\label{sec5}

This work proposes Mobile-GRAPPA as a practical way to reconcile dense motion and $\delta B_0$ tracking with feasible reconstruction times. Instead of embedding every motion state directly into the forward operator, Mobile-GRAPPA operates in k-space: it learns local GRAPPA operators that map motion- and $\delta B_0$-corrupted data to a common reference state, after which reconstruction proceeds with standard SENSE or any advanced pipeline of choice. In this way, motion and $\delta B_0$ modeling is decoupled from the image-space solve.

Across detailed characterizations and generalized applications, Mobile-GRAPPA consistently achieved artifact suppression comparable to full Aligned-SENSE while remaining computationally efficient. It reduces ringing, ghosting, and blurring in heavily motion-corrupted 3D multi-echo GRE at 3T, stabilized EPTI under both heavy motion and deep breathing, and remained robust at 7T where $\delta B_0$ and coil variation are stronger. Relative to Approx-Augmented-SENSE (trajectory rotation and linear phase only), Mobile-GRAPPA adds learned corrections for coil reweighting and $\delta B_0$ phase. 

Mobile-GRAPPA also changes the computational scaling behavior of motion and $\delta B_0$ modeling. In Aligned-SENSE and Augmented-SENSE, cost grows roughly linearly with the number of motion states because each state contributes a distinct encoding operator that must be applied in every reconstruction iteration. In contrast, Mobile-GRAPPA adds a small, nearly constant overhead: brief MLP training (seconds per volume) and a $\sim$1 s k-space application step, after which reconstruction proceeds at the no-motion baseline complexity. When clustering is triggered for very large motions, runtime scales with the number of clusters rather than the number of states, and the required number of clusters in practice is modest. Among all the experiments, we at most needed 2 clusters to achieve comparable result to Aligned-SENSE. 

An additional benefit of Mobile-GRAPPA is that it preserves rather than compresses temporal information. Because all states are used during training and application, tractability is achieved without sacrificing motion and $\delta B_0$ temporal resolution.

The MLP-based representation of a kernel family is not limited to motion correction. In this work, Mobile-GRAPPA is shown to be well suited for correcting pose-dependent and breathing-related $\delta B_0$ changes, which are typically spatially smooth and thus amenable to local-kernel correction. More broadly, this formulation provides a mechanism for representing spatially smooth image-domain imperfections as local k-space operators. For example, eddy-current-induced phase, which is often spatially smooth, may also be amenable to correction using learned local kernels. In contrast, Mobile-GRAPPA should not be expected to directly correct the full static $B_0$ inhomogeneity when the associated phase becomes too spatially complex, particularly at long TE, because such structure may not be represented stably by strictly local k-space operators. 

Mobile-GRAPPA is naturally compatible with a wide range of reconstruction strategies. Because the cleaned k-space data resemble acquisition in a single reference state, they can be used with conventional CG-SENSE, spatiotemporal subspace reconstructions such as EPTI, or formulations incorporating learned priors and plug-in regularizes. In each case, the data-consistency operator remains unchanged because Mobile-GRAPPA modifies only the input k-space data.

If the motion and $\delta B_0$ estimations are inaccurate, the Mobile-GRAPPA-based cleaning step will leave residual inconsistencies in k-space. Prior work \cite{hamilton2017parallelImagingReview} suggests that GRAPPA-style parallel imaging reconstructions can be more robust than SENSE under model mismatch from motions because GRAPPA performs local k-space operators on data acquired with limited motion, whereas SENSE enforces global unaliasing that can amplify structured errors when the forward model is wrong. This observation suggests a natural potential extension of our work for imaging cases where motion and $\delta B_0$ information are available but are of limited accuracy. In such scenario, Mobile-GRAPPA can be applied to partially cleaned the motion-corrupted data, which can then be subsequently reconstructed using  GRAPPA-based image reconstruction.

Other potential avenues of future work includes integrating Mobile-GRAPPA into joint image–motion estimation frameworks, where motion and $\delta B_0$ estimates are refined iteratively from reconstructed images. Another direction is to extend beyond rigid-body motion by modeling spatially varying motion as locally rigid and augmenting the MLP features of Mobile-GRAPPA to represent non-rigid deformation.

\section{Conclusions}\label{sec6}

Mobile-GRAPPA is a motion- and $\delta B_0$-aware k-space cleaning framework that uses an MLP-parameterized family of local GRAPPA kernels to map motion- and $\delta B_0$-corrupted k-space data to a desired target state representation, either directly to a single reference state or, in extreme motion regimes, through a small number of intermediate clustered states. This one-time operation performs effective trajectory gridding, coil reweighting correction, and $\delta B_0$ phase correction, enabling the incorporation of dense motion and $\delta B_0$ modeling with a negligible increase in reconstruction time in the standard setting and modest additional cost when clustering is used. The cleaned k-space data can be reconstructed using standard SENSE, subspace methods, or other pipelines without modifying their forward operators. Across detailed characterizations and in vivo experiments at 3T and 7T, Mobile-GRAPPA achieved image quality comparable to Aligned-SENSE without significant noise amplification and with reconstruction times close to no-motion baselines.

\subsection*{Funding}

This work was supported by Siemens Healthineers and NIH research
grants: R01MH116173; R01EB019437; R01HD114719; R21EB038677.

\subsection*{Data Availability Statement}

The reconstruction code and example data for the Mobile-GRAPPA pipeline are publicly available at \url{https://github.com/linym20/Mobile-GRAPPA}.

\bibliography{MRM-AMA}%

@article{zaitsev2015motion,
  title   = {Motion artifacts in {MRI}: A complex problem with many partial solutions},
  author  = {Zaitsev, Maxim and Maclaren, Julian and Herbst, Michael},
  journal = {J Magn Reson Imaging},
  year    = {2015},
  volume  = {42},
  number  = {4},
  pages   = {887--901},
  doi     = {10.1002/jmri.24850}
}

@article{godenschweger2016brainmoco,
  title   = {Motion correction in {MRI} of the brain},
  author  = {Godenschweger, Frank and K{\"a}gebein, Urte and Stucht, Daniel and Yarach, Uten and Sciarra, Alessandro and Yakupov, Renat and L{\"u}sebrink, Falk and Schulze, Peter and Speck, Oliver},
  journal = {Phys Med Biol},
  year    = {2016},
  volume  = {61},
  number  = {5},
  pages   = {R32--R56},
  doi     = {10.1088/0031-9155/61/5/R32}
}

@article{bammer2007agSense,
  title   = {Augmented generalized {SENSE} reconstruction to correct for rigid body motion},
  author  = {Bammer, Roland and Aksoy, Murat and Liu, Chunlei},
  journal = {Magn Reson Med},
  year    = {2007},
  volume  = {57},
  number  = {1},
  pages   = {90--102},
  doi     = {10.1002/mrm.21106}
}

@article{farajidana2016coilMotion,
  title   = {Interactions between head motion and coil sensitivity in accelerated {fMRI}},
  author  = {Faraji-Dana, Zahra and Tam, Fred and Chen, Jean J. and Graham, Simon J.},
  journal = {J Neurosci Methods},
  year    = {2016},
  volume  = {270},
  pages   = {46--60},
  doi     = {10.1016/j.jneumeth.2016.06.005}
}

@article{liu2018headmotionB0,
  title   = {Effect of head motion on {MRI} {$B_0$} field distribution},
  author  = {Liu, Jiaen and de Zwart, Jacco A. and van Gelderen, Peter and Murphy-Boesch, Joseph and Duyn, Jeff H.},
  journal = {Magn Reson Med},
  year    = {2018},
  volume  = {80},
  number  = {6},
  pages   = {2538--2548},
  doi     = {10.1002/mrm.27339}
}

@article{brackenier2022posedependentB0,
  title   = {Data-driven motion-corrected brain {MRI} incorporating pose-dependent {$B_0$} fields},
  author  = {Brackenier, Yannick and Cordero-Grande, Lucilio and Tomi-Tricot, Raphael and Wilkinson, Thomas and Bridgen, Philippa and Price, Anthony and Malik, Shaihan J. and De Vita, Enrico and Hajnal, Joseph V.},
  journal = {Magn Reson Med},
  year    = {2022},
  volume  = {88},
  number  = {2},
  pages   = {817--831},
  doi     = {10.1002/mrm.29255}
}

@article{laustsen2022eegmotion,
  title   = {Tracking of rigid head motion during {MRI} using an {EEG} system},
  author  = {Laustsen, Malte and Andersen, Mads and Xue, Rong and Madsen, Kristoffer H. and Hanson, Lars G.},
  journal = {Magn Reson Med},
  year    = {2022},
  volume  = {88},
  number  = {2},
  pages   = {986--1001},
  doi     = {10.1002/mrm.29251}
}

@article{vanniekerk2019plugplay,
  title   = {Toward ``plug and play'' prospective motion correction for {MRI} by combining observations of the time varying gradient and static vector fields},
  author  = {van Niekerk, Adam and van der Kouwe, Andre and Meintjes, Ernesta},
  journal = {Magn Reson Med},
  year    = {2019},
  volume  = {82},
  number  = {3},
  pages   = {1214--1228},
  doi     = {10.1002/mrm.27790}
}

@article{maclaren2012micromotion,
  title   = {Measurement and correction of microscopic head motion during magnetic resonance imaging of the brain},
  author  = {Maclaren, Julian and Armstrong, Brian S. R. and Barrows, Robert T. and Danishad, K. Appu and Ernst, Theodor and Foster, Colin L. and G{\"u}m{\"u}\c{s}, Kazim and Herbst, Michael and Kadashevich, Ilja Y. and Kusik, Todd P. and Li, Qiaotian and Lovell-Smith, Christopher and Prieto, Claudia and Schulze, Patrick and Speck, Oliver and Stucht, Daniel and Zaitsev, Maxim},
  journal = {PLoS One},
  year    = {2012},
  volume  = {7},
  number  = {11},
  pages   = {e48088},
  doi     = {10.1371/journal.pone.0048088}
}

@article{gretsch2020fatnavsMPT,
  title   = {Fat navigators and {Moir{\'e}} phase tracking comparison for motion estimation and retrospective correction},
  author  = {Gretsch, Fr{\'e}d{\'e}ric and Mattern, Hendrik and Gallichan, Daniel and Speck, Oliver},
  journal = {Magn Reson Med},
  year    = {2020},
  volume  = {83},
  number  = {1},
  pages   = {83--93},
  doi     = {10.1002/mrm.27908}
}

@article{jorge2018trackdots,
  title   = {Tracking discrete off-resonance markers with three spokes (trackDOTS) for compensation of head motion and {$B_0$} perturbations: Accuracy and performance in anatomical imaging},
  author  = {Jorge, Jo{\~a}o and Gretsch, Fr{\'e}d{\'e}ric and Gallichan, Daniel and Marques, Jos{\'e} P.},
  journal = {Magn Reson Med},
  year    = {2018},
  volume  = {79},
  number  = {1},
  pages   = {160--171},
  doi     = {10.1002/mrm.26654}
}

@article{ulrich2024servonavigators,
  title   = {Servo navigators: linear regression and feedback control for rigid-body motion correction},
  author  = {Ulrich, Thomas and Riedel, Malte and Pruessmann, Klaas P.},
  journal = {Magn Reson Med},
  year    = {2024},
  volume  = {91},
  number  = {5},
  pages   = {1876--1892},
  doi     = {10.1002/mrm.29967}
}

@article{brackenier2024queen,
  title   = {Rapid and accurate navigators for motion and {$B_0$} tracking using {QUEEN}: Quantitatively enhanced parameter estimation from navigators},
  author  = {Brackenier, Yannick and Wang, Nan and Liao, Congyu and Cao, Xiaozhi and Schauman, Sophie and Yurt, Mahmut and Cordero-Grande, Lucilio and Malik, Shaihan J. and Kerr, Adam and Hajnal, Joseph V. and Setsompop, Kawin},
  journal = {Magn Reson Med},
  year    = {2024},
  volume  = {91},
  number  = {5},
  pages   = {2028--2043},
  doi     = {10.1002/mrm.29976}
}

@article{polak2022samer,
  title   = {Scout accelerated motion estimation and reduction ({SAMER})},
  author  = {Polak, Daniel and Splitthoff, Daniel Nicolas and Clifford, Bryan and Lo, Wei-Ching and Huang, Susie Y. and Conklin, John and Wald, Lawrence L. and Setsompop, Kawin and Cauley, Stephen},
  journal = {Magn Reson Med},
  year    = {2022},
  volume  = {87},
  number  = {1},
  pages   = {163--178},
  doi     = {10.1002/mrm.28971}
}

@article{corderogrande2016alignedMultishot,
  title   = {Sensitivity encoding for aligned multishot magnetic resonance reconstruction},
  author  = {Cordero-Grande, Lucilio and Teixeira, Rui Pedro A. G. and Hughes, Emer J. and Hutter, Jana and Price, Anthony N. and Hajnal, Joseph V.},
  journal = {IEEE Trans Comput Imaging},
  year    = {2016},
  volume  = {2},
  number  = {3},
  pages   = {266--280},
  doi     = {10.1109/TCI.2016.2557069}
}

@article{liu2020reducingMotionSensitivity,
  title   = {Reducing motion sensitivity in 3D high-resolution {T$_2$}*-weighted {MRI} by navigator-based motion and nonlinear magnetic field correction},
  author  = {Liu, Jiaen and van Gelderen, Peter and de Zwart, Jacco A. and Duyn, Jeff H.},
  journal = {Neuroimage},
  year    = {2020},
  volume  = {206},
  pages   = {116332},
  doi     = {10.1016/j.neuroimage.2019.116332}
}

@article{yarach2016distortionVar,
  title   = {Correction of {$B_0$}-induced geometric distortion variations in prospective motion correction for 7{T} {MRI}},
  author  = {Yarach, Uten and Luengviriya, Chaiya and Stucht, Daniel and Godenschweger, Frank and Schulze, Peter and Speck, Oliver},
  journal = {MAGMA},
  year    = {2016},
  volume  = {29},
  number  = {3},
  pages   = {319--332},
  doi     = {10.1007/s10334-015-0515-2}
}

@article{wang2019epti,
  title   = {Echo planar time-resolved imaging ({EPTI})},
  author  = {Wang, Fuyixue and Dong, Zijing and Reese, Timothy G. and Bilgic, Berkin and Manhard, Mary Katherine and Chen, Jingyuan and Polimeni, Jonathan R. and Wald, Lawrence L. and Setsompop, Kawin},
  journal = {Magn Reson Med},
  year    = {2019},
  volume  = {81},
  number  = {6},
  pages   = {3599--3615},
  doi     = {10.1002/mrm.27673}
}

@article{dong2020eptiSubspace,
  title   = {Echo planar time-resolved imaging with subspace reconstruction and optimized spatiotemporal encoding},
  author  = {Dong, Zijing and Wang, Fuyixue and Reese, Timothy G. and Bilgic, Berkin and Setsompop, Kawin},
  journal = {Magn Reson Med},
  year    = {2020},
  volume  = {84},
  number  = {5},
  pages   = {2442--2455},
  doi     = {10.1002/mrm.28295}
}

@article{dong2022motionCorrected3dEPTI,
  title   = {Motion-corrected 3D-{EPTI} with efficient 4D navigator acquisition for fast and robust whole-brain quantitative imaging},
  author  = {Dong, Zijing and Wang, Fuyixue and Setsompop, Kawin},
  journal = {Magn Reson Med},
  year    = {2022},
  volume  = {88},
  number  = {3},
  pages   = {1112--1125},
  doi     = {10.1002/mrm.29277}
}

@article{xu2019mrfSlidingWindow,
  title   = {Rigid motion correction for magnetic resonance fingerprinting with sliding-window reconstruction and image registration},
  author  = {Xu, Zhongbiao and Ye, Huihui and Lyu, Mengye and He, Hongjian and Zhong, Jianhui and Mei, Yingjie and Chen, Zhifeng and Wu, Ed X. and Chen, Wufan and Feng, Qianjin and Feng, Yanqiu},
  journal = {Magn Reson Imaging},
  year    = {2019},
  volume  = {57},
  pages   = {303--312},
  doi     = {10.1016/j.mri.2018.11.001}
}

@article{cruz2019mcmrf,
  title   = {Rigid motion-corrected magnetic resonance fingerprinting},
  author  = {Cruz, Gast{\~a}o and Jaubert, Olivier and Schneider, Torben and Botnar, Rene M. and Prieto, Claudia},
  journal = {Magn Reson Med},
  year    = {2019},
  volume  = {81},
  number  = {2},
  pages   = {947--961},
  doi     = {10.1002/mrm.27448}
}

@article{griswold2002grappa,
  title   = {Generalized autocalibrating partially parallel acquisitions ({GRAPPA})},
  author  = {Griswold, Mark A. and Jakob, Peter M. and Heidemann, Robin M. and Nittka, Mathias and Jellus, Vladimir and Wang, Jianmin and Kiefer, Berthold and Haase, Axel},
  journal = {Magn Reson Med},
  year    = {2002},
  volume  = {47},
  number  = {6},
  pages   = {1202--1210},
  doi     = {10.1002/mrm.10171}
}

@article{hoge2016dpgrappa,
  title   = {Dual-polarity {GRAPPA} for simultaneous reconstruction and ghost correction of echo planar imaging data},
  author  = {Hoge, W. Scott and Polimeni, Jonathan R.},
  journal = {Magn Reson Med},
  year    = {2016},
  volume  = {76},
  number  = {1},
  pages   = {32--44},
  doi     = {10.1002/mrm.25839}
}

@article{abraham2023implicitgrappa,
  title         = {Implicit Representation of {GRAPPA} Kernels for Fast {MRI} Reconstruction},
  author        = {Abraham, Daniel Raz and Nishimura, Mark and Cao, Xiaozhi and Liao, Congyu and Setsompop, Kawin},
  journal       = {arXiv},
  year          = {2023},
  eprint        = {2310.10823},
  archivePrefix = {arXiv},
  doi           = {10.48550/arXiv.2310.10823},
  url           = {https://arxiv.org/abs/2310.10823}
}

@article{stockmann2018shimming,
  title   = {In vivo {$B_0$} field shimming methods for {MRI} at 7 {T}},
  author  = {Stockmann, Jason P. and Wald, Lawrence L.},
  journal = {Neuroimage},
  year    = {2018},
  volume  = {168},
  pages   = {71--87},
  doi     = {10.1016/j.neuroimage.2017.06.013}
}

@article{robson2008snrgfactor,
  title   = {Comprehensive quantification of signal-to-noise ratio and g-factor for image-based and k-space-based parallel imaging reconstructions},
  author  = {Robson, Philip M. and Grant, Aaron K. and Madhuranthakam, Ananth J. and Lattanzi, Riccardo and Sodickson, Daniel K. and McKenzie, Charles A.},
  journal = {Magn Reson Med},
  year    = {2008},
  volume  = {60},
  number  = {4},
  pages   = {895--907},
  doi     = {10.1002/mrm.21728}
}

@article{langkammer2012qsmbrainiron,
  title   = {Quantitative susceptibility mapping ({QSM}) as a means to measure brain iron? A post mortem validation study},
  author  = {Langkammer, Christian and Schweser, Ferdinand and Krebs, Nikolaus and Deistung, Andreas and Goessler, Walter and Scheurer, Eva and Sommer, Karsten and Reishofer, Gernot and Yen, Kathrin and Fazekas, Franz and Ropele, Stefan and Reichenbach, J{\"u}rgen R.},
  journal = {Neuroimage},
  year    = {2012},
  volume  = {62},
  number  = {3},
  pages   = {1593--1599},
  doi     = {10.1016/j.neuroimage.2012.05.049}
}

@article{hamilton2017parallelImagingReview,
  title   = {Recent advances in parallel imaging for {MRI}},
  author  = {Hamilton, Jesse and Franson, Dominique and Seiberlich, Nicole},
  journal = {Prog Nucl Magn Reson Spectrosc},
  year    = {2017},
  volume  = {101},
  pages   = {71--95},
  doi     = {10.1016/j.pnmrs.2017.04.002}
}

@article{meng2025qsmB0,
  title   = {Motion and temporal {$B_0$}-shift corrections for {QSM} and {$R_2^\ast$} mapping using dual-echo spiral navigators and conjugate-phase reconstruction},
  author  = {Meng, Yuguang and Allen, Jason W. and Sharghi, Vahid Khalilzad and Qiu, Deqiang},
  journal = {Magn Reson Med},
  year    = {2025},
  volume  = {93},
  number  = {1},
  pages   = {199--212},
  doi     = {10.1002/mrm.30266}
}

@article{wiesinger2004piFieldStrength,
  title   = {Parallel imaging performance as a function of field strength---an experimental investigation using electrodynamic scaling},
  author  = {Wiesinger, Florian and van de Moortele, Pierre-Francois and Adriany, Gregor and de Zanche, Nicola and Ugurbil, Kamil and Pruessmann, Klaas P.},
  journal = {Magn Reson Med},
  year    = {2004},
  volume  = {52},
  number  = {5},
  pages   = {953--964},
  doi     = {10.1002/mrm.20281}
}

@article{seiberlich2007grog,
  title   = {Non-Cartesian data reconstruction using {GRAPPA} operator gridding ({GROG})},
  author  = {Seiberlich, Nicole and Breuer, Felix A. and Blaimer, Martin and Barkauskas, Kestutis and Jakob, Peter M. and Griswold, Mark A.},
  journal = {Magn Reson Med},
  year    = {2007},
  volume  = {58},
  number  = {6},
  pages   = {1257--1265},
  doi     = {10.1002/mrm.21435}
}

@article{breuer2005caipirinha,
  title   = {Controlled aliasing in parallel imaging results in higher acceleration ({CAIPIRINHA}) for multi-slice imaging},
  author  = {Breuer, Felix A. and Blaimer, Martin and Heidemann, Robin M. and Mueller, Matthias F. and Griswold, Mark A. and Jakob, Peter M.},
  journal = {Magn Reson Med},
  year    = {2005},
  volume  = {53},
  number  = {3},
  pages   = {684--691},
  doi     = {10.1002/mrm.20401}
}

@article{wang2025septi,
  title   = {Spherical echo-planar time-resolved imaging ({sEPTI}) for rapid 3D quantitative {T2*} and susceptibility imaging},
  author  = {Wang, Nan and Liao, Congyu and Cao, Xiaozhi and Nishimura, Mark and Brackenier, Yannick and Yurt, Mahmut and Gao, Mengze and Abraham, Daniel Raz and Alkan, Cagan and Iyer, Siddharth Srinivasan and Zhou, Zihan and Jeong, Hwihun and Kerr, Adam and Haldar, Justin P. and Setsompop, Kawin},
  journal = {Magn Reson Med},
  year    = {2025},
  volume  = {93},
  number  = {1},
  pages   = {121--137},
  doi     = {10.1002/mrm.30255}
}

@article{berglund2021markerlessdwi,
  title   = {Prospective motion correction for diffusion weighted {EPI} of the brain using an optical markerless tracker},
  author  = {Berglund, Johan and van Niekerk, Adam and Ryd{\'e}n, Henric and Sprenger, Tim and Avventi, Enrico and Norbeck, Ola and Glimberg, Stefan L. and Olesen, Oline V. and Skare, Stefan},
  journal = {Magn Reson Med},
  year    = {2021},
  volume  = {85},
  number  = {3},
  pages   = {1427--1440},
  doi     = {10.1002/mrm.28524}
}

@article{chen2023multishotmarkerless,
  title   = {High-resolution multi-shot diffusion-weighted {MRI} combining markerless prospective motion correction and locally low-rank constrained reconstruction},
  author  = {Chen, Hao and Dai, Ke and Zhong, Sijie and Zheng, Jiaxu and Zhang, Xinyue and Yang, Shasha and Cao, Tuoyu and Wang, Chaohong and Karasan, Ekin and Frydman, Lucio and Zhang, Zhiyong},
  journal = {Magn Reson Med},
  year    = {2023},
  volume  = {89},
  number  = {2},
  pages   = {605--619},
  doi     = {10.1002/mrm.29468}
}

@article{brackenier2024pilottonedmc,
  title   = {Sequence-agnostic motion-correction leveraging efficiently calibrated {Pilot Tone} signals},
  author  = {Brackenier, Yannick and Cordero-Grande, Lucilio and McElroy, Sarah and Tomi-Tricot, Raphael and Barbaroux, Hugo and Bridgen, Philippa and Malik, Shaihan J. and Hajnal, Joseph V.},
  journal = {Magn Reson Med},
  year    = {2024},
  volume  = {92},
  number  = {5},
  pages   = {1881--1897},
  doi     = {10.1002/mrm.30161}
}

@article{anand2024beatpilottone,
  title   = {Beat {Pilot Tone} ({BPT}): Simultaneous {MRI} and {RF} motion sensing at arbitrary frequencies},
  author  = {Anand, Suma and Lustig, Michael},
  journal = {Magn Reson Med},
  year    = {2024},
  volume  = {92},
  number  = {4},
  pages   = {1768--1787},
  doi     = {10.1002/mrm.30150}
}

@article{tisdall2012vnavigators,
  title   = {Volumetric navigators for prospective motion correction and selective reacquisition in neuroanatomical {MRI}},
  author  = {Tisdall, M. Dylan and Hess, Aaron T. and Reuter, Martin and Meintjes, Ernesta M. and Fischl, Bruce and van der Kouwe, Andr{\'e} J. W.},
  journal = {Magn Reson Med},
  year    = {2012},
  volume  = {68},
  number  = {2},
  pages   = {389--399},
  doi     = {10.1002/mrm.23228}
}

@article{vanderkouwe2006cloverleaf,
  title   = {Real-time rigid body motion correction and shimming using cloverleaf navigators},
  author  = {van der Kouwe, Andr{\'e} J. W. and Benner, Thomas and Dale, Anders M.},
  journal = {Magn Reson Med},
  year    = {2006},
  volume  = {56},
  number  = {5},
  pages   = {1019--1032},
  doi     = {10.1002/mrm.21038}
}

@article{white2010promo,
  title   = {{PROMO}: real-time prospective motion correction in {MRI} using image-based tracking},
  author  = {White, Nathan and Roddey, Cooper and Shankaranarayanan, Ajit and Han, Eric and Rettmann, Dan and Santos, Juan and Kuperman, Josh and Dale, Anders},
  journal = {Magn Reson Med},
  year    = {2010},
  volume  = {63},
  number  = {1},
  pages   = {91--105},
  doi     = {10.1002/mrm.22176}
}

@article{wallace2019fidnavs,
  title   = {Head motion measurement and correction using {FID} navigators},
  author  = {Wallace, Tess E. and Afacan, Onur and Waszak, Maryna and Kober, Tobias and Warfield, Simon K.},
  journal = {Magn Reson Med},
  year    = {2019},
  volume  = {81},
  number  = {1},
  pages   = {258--274},
  doi     = {10.1002/mrm.27381}
}

@article{gallichan2016fatnavs7t,
  title   = {Retrospective correction of involuntary microscopic head movement using highly accelerated fat image navigators ({3D FatNavs}) at 7{T}},
  author  = {Gallichan, Daniel and Marques, Jos{\'e} P. and Gruetter, Rolf},
  journal = {Magn Reson Med},
  year    = {2016},
  volume  = {75},
  number  = {3},
  pages   = {1030--1039},
  doi     = {10.1002/mrm.25670}
}

@article{zaitsev2006opticalpmc,
  title   = {Magnetic resonance imaging of freely moving objects: prospective real-time motion correction using an external optical motion tracking system},
  author  = {Zaitsev, Maxim and Dold, Christian and Sakas, Georg and Hennig, J{\"u}rgen and Speck, Oliver},
  journal = {Neuroimage},
  year    = {2006},
  volume  = {31},
  number  = {3},
  pages   = {1038--1050},
  doi     = {10.1016/j.neuroimage.2006.01.039}
}

@article{stucht2015highestrespmc,
  title   = {Highest resolution in vivo human brain {MRI} using prospective motion correction},
  author  = {Stucht, Daniel and Danishad, K. Appu and Schulze, Peter and Godenschweger, Frank and Zaitsev, Maxim and Speck, Oliver},
  journal = {PLoS One},
  year    = {2015},
  volume  = {10},
  number  = {7},
  pages   = {e0133921},
  doi     = {10.1371/journal.pone.0133921}
}

@article{vanniekerk2019wrad,
  title   = {A wireless radio frequency triggered acquisition device ({WRAD}) for self-synchronised measurements of the rate of change of the {MRI} gradient vector field for motion tracking},
  author  = {van Niekerk, Adam and Meintjes, Ernesta and van der Kouwe, Andre},
  journal = {IEEE Trans Med Imaging},
  year    = {2019},
  volume  = {38},
  number  = {7},
  pages   = {1610--1621},
  doi     = {10.1109/TMI.2019.2891774}
}

@article{maclaren2013pmcreview,
  title   = {Prospective motion correction in brain imaging: a review},
  author  = {Maclaren, Julian and Herbst, Michael and Speck, Oliver and Zaitsev, Maxim},
  journal = {Magn Reson Med},
  year    = {2013},
  volume  = {69},
  number  = {3},
  pages   = {621--636},
  doi     = {10.1002/mrm.24314}
}

@article{vionnet2021jointfieldmotion,
  title   = {Simultaneous feedback control for joint field and motion correction in brain MRI},
  author  = {Vionnet, Laetitia and Aranovitch, Alexander and Duerst, Yolanda and Haeberlin, Maximilian and Dietrich, Benjamin Emmanuel and Gross, Simon and Pruessmann, Klaas Paul},
  journal = {Neuroimage},
  year    = {2021},
  volume  = {226},
  pages   = {117286},
  doi     = {10.1016/j.neuroimage.2020.117286}
}

@article{ludwig2021pilottonecine,
  title   = {Pilot tone-based motion correction for prospective respiratory compensated cardiac cine MRI},
  author  = {Ludwig, Juliane and Speier, Peter and Seifert, Frank and Schaeffter, Tobias and Kolbitsch, Christoph},
  journal = {Magn Reson Med},
  year    = {2021},
  volume  = {85},
  number  = {5},
  pages   = {2403--2416},
  doi     = {10.1002/mrm.28580}
}

@article{speier2015ptnav,
  title   = {{PT-nav}: a novel respiratory navigation method for continuous acquisition based on modulation of a {Pilot Tone} on the {MR}-receiver},
  author  = {Speier, Peter and Fenchel, Michael and Rehner, Ralf},
  journal = {Magn Reson Mater Phy},
  year    = {2015},
  volume  = {28},
  pages   = {S97--S98},
  doi     = {10.1007/s10334-015-0487-2}
}

@inproceedings{aksoy2008coilSensitivityPI,
  title   = {Effect of Motion-Induced Altered Coil Sensitivity on Parallel Imaging Performance},
  author  = {Aksoy, Murat and Bammer, Roland},
  series  = {Proceedings of the 16th Annual Meeting of the {ISMRM}},
  address = {Toronto, ON, Canada},
  year    = {2008},
  note    = {Abstract 3111}
}

@inproceedings{luengviriya2010sensProfile,
  title   = {Necessity of sensitivity profile correction in retrospective motion correction},
  author  = {Luengviriya, Chaiya and Yun, Jian and Lee, Kuan and Maclaren, Julian and Speck, Oliver},
  series  = {Proceedings of the 18th Annual Meeting of the {ISMRM}},
  address = {Stockholm, Sweden},
  year    = {2010},
  note    = {Abstract 3064}
}

@inproceedings{wang2025smena,
  title   = {Scout-based Multi-Echo NAvigating ({SMENA}) for high temporal resolution motion and {$B_0$} estimation: applications to {EPTI} and multi-echo {GRE}},
  author  = {Wang, Nan and Brackenier, Yannick W. E. and Nurdinova, Aizada and Zhou, Zihan and Abraham, Daniel Raz and Lin, Yimeng and Cao, Xiaozhi and Liao, Congyu and Setsompop, Kawin},
  series  = {Proceedings of the Honolulu - 2025 {ISMRM} Annual Meeting},
  address = {Honolulu, HI, USA},
  year    = {2025},
  note    = {Abstract 0034},
  url     = {https://archive.ismrm.org/2025/0034.html}
}

@inproceedings{wang2026stsmena,
  author = {Wang, Nan and Lin, Yimeng and Polak, Daniel and Cao, Xiaozhi and Urman, Yonatan and Nurdinova, Aizada and Gao, Mengze and Abraham, Daniel Raz and Shah, Zachary Andrew and Liao, Congyu and Cauley, Stephen and Setsompop, Kawin},
  title  = {Spatiotemporal Scout-based Multi-Echo NAvigator (st-SMENA) for Accurate and Continuous Motion and {$\delta B_0$} Tracking},
  series  = {Proceedings of the Cape Town - 2026 ISMRM-ISMRT Annual Meeting and Exhibition},
  address = {Cape Town, South Africa},
  year    = {2026},
  note    = {Program \#560-01-004}
}

@inproceedings{wang2024fcg,
  title   = {Field-Correcting {GRAPPA} ({FCG}) for improved mitigation of even-odd and field-related artifacts in {EPI}},
  author  = {Wang, Nan and Abraham, Daniel Raz and Kerr, Adam B. and Wu, Hua and Liao, Congyu and Cao, Xiaozhi and Polimeni, Jonathan R. and Huber, Renzo and Setsompop, Kawin},
  series  = {Proceedings of the 2024 {ISMRM} Annual Meeting},
  address = {Singapore},
  year    = {2024},
  note    = {Abstract 1256},
  url     = {https://archive.ismrm.org/2024/1256.html}
}

@inproceedings{macqueen1967classification,
  title     = {Some methods for classification and analysis of multivariate observations},
  author    = {MacQueen, James B.},
  series    = {Proceedings of the Fifth Berkeley Symposium on Mathematical Statistics and Probability, Volume 1: Statistics},
  publisher = {University of California Press},
  address   = {Berkeley, CA},
  year      = {1967},
  pages     = {281--297}
}

@inproceedings{arthur2006kmeanspp,
  title     = {k-means++: The advantages of careful seeding},
  author    = {Arthur, David and Vassilvitskii, Sergei},
  series    = {Proceedings of the Eighteenth Annual {ACM}-{SIAM} Symposium on Discrete Algorithms},
  address   = {New Orleans, LA},
  year      = {2007},
  pages     = {1027--1035},
  doi       = {10.5555/1283383.1283494}
}

@inproceedings{liang2007spatiotemporal,
  title   = {Spatiotemporal Imaging with Partially Separable Functions},
  author  = {Liang, Zhi-Pei},
  series  = {Proc. of 2007 Joint Meet. of the 6th Int. Symp. on Noninvasive Functional Source Imaging of the Brain and Heart and the Int. Conf. on Functional Biomedical Imaging, {NFSI} and {ICFBI} 2007},
  address = {Hangzhou, China},
  year    = {2007},
  pages   = {181--182},
  doi     = {10.1109/NFSI-ICFBI.2007.4387720}
}

@inproceedings{vanniekerk2024shortTRpmc,
  title   = {Navigator based prospective motion correction in short {TR} sequences with minimal scan time penalty},
  author  = {van Niekerk, Adam and Ryd{\'e}n, Henric and Schauman, Sophie and Norbeck, Ola and Sprenger, Tim and Avventi, Enrico and Skare, Stefan},
  series  = {Proceedings of the 2024 {ISMRM} Annual Meeting},
  address = {Singapore},
  year    = {2024},
  note    = {Abstract 0388},
  url     = {https://archive.ismrm.org/2024/0388.html}
}

@inproceedings{wilkinson2021pilottoneuhf,
  title   = {Motion Estimation for Brain Imaging at Ultra-High Field Using Pilot-Tone: Comparison with {DISORDER} Motion Compensation},
  author  = {Wilkinson, Tom and Godinez, Felipe and Brackenier, Yannick and Tomi-Tricot, Raphael and Cordero-Grande, Lucilio and Bridgen, Philippa and Giles, Sharon and Hajnal, Joseph V. and Malik, Shaihan J.},
  series  = {Proceedings of the 29th Annual Meeting of the {ISMRM}},
  year    = {2021},
  note    = {Abstract 0122},
  url     = {https://archive.ismrm.org/2021/0122.html}
}

@inproceedings{yarach2015coilSensitivityMiscalibration,
  title   = {The Correction of Motion-Induced Coil Sensitivity Miscalibration in Parallel Imaging with Prospective Motion Correction},
  author  = {Yarach, Uten and Stucht, Daniel and Godenschweger, Frank and Speck, Oliver},
  series  = {Proceedings of the 23rd Annual Meeting of the {ISMRM}},
  year    = {2015},
  note    = {Abstract 2558},
  url     = {https://archive.ismrm.org/2015/2558.html}
}

@inproceedings{lin2026mobilegrappa,
  title   = {Fast Reconstruction of Motion-Corrupted Data with Mobile-GRAPPA: Motion and dB0 Correction Leveraging Efficient GRAPPA},
  author  = {Lin, Yimeng and Wang, Nan and Abraham, Daniel Raz and Polak, Daniel and Cao, Xiaozhi and Cauley, Stephen and Setsompop, Kawin},
  series  = {Proceedings of the Cape Town - 2026 ISMRM-ISMRT Annual Meeting and Exhibition},
  address = {Cape Town, South Africa},
  year    = {2026},
  note    = {Program \#605-02-002}
}

@inproceedings{lin2025motionigrog,
  title   = {Fast and accurate motion-corrected reconstruction with motion-correcting Implicit GROG (motion-iGROG)},
  author  = {Lin, Yimeng and Abraham, Daniel Raz and Wang, Nan and Zhou, Zihan and Cao, Xiaozhi and Nurdinova, Aizada and Setsompop, Kawin},
  series  = {Proceedings of the 2025 ISMRM and ISMRT Annual Meeting and Exhibition},
  address = {Honolulu, HI, United States of America},
  year    = {2025},
  note    = {Abstract 4434},
  url     = {https://archive.ismrm.org/2025/4434.html}
}

\end{document}